\documentclass[fleqn,usenatbib,useAMS]{mnras}

\usepackage{graphicx}	
\usepackage{amsmath}	
\usepackage{amssymb}	
\usepackage{multicol}        
\usepackage{bm}		
\usepackage{pdflscape}	

\usepackage[T1]{fontenc}

\DeclareRobustCommand{\VAN}[3]{#2}
\let\VANthebibliography\thebibliography
\def\thebibliography{\DeclareRobustCommand{\VAN}[3]{##3}\VANthebibliography}

\usepackage{graphicx}	
\usepackage{amsmath}	

\newcommand{\ketju}[0]{\textsc{ketju}}

\usepackage{color}

\graphicspath{{figures/}}

\title[High redshift galaxies and SMBH binaries]{Rapid sinking and efficient mergers of supermassive black holes in compact high-redshift galaxies}

\author[A. Keitaanranta et al.]
{Atte Keitaanranta,$^{1}$\thanks{Email: atte.keitaanranta@helsinki.fi} Peter H. Johansson,$^{1}$ Alexander Rawlings,$^{1}$ Toni Tuominen,$^{1}$ Antti Rantala,$^{2,3,4}$ \newauthor Thorsten Naab,$^{4}$  Shihong Liao,$^{5}$ Bastián Reinoso$^{1}$ \vspace{0.2cm}
\\
$^1$Department of Physics, University of Helsinki, Gustaf Hällströmin katu 2, FI-00014 Helsinki, Finland\\
$^{2}$Institute of Astronomy, University of Cambridge, Madingley Road, Cambridge CB3 0HA, UK\\
$^{3}$Kavli Institute for Cosmology, Cambridge (KICC), University of Cambridge, Madingley Road, Cambridge CB3 0HA, UK\\
$^4$Max-Planck-Institut f\"ur Astrophysik, Karl-Schwarzchild-Str 1, D-85748 Garching, Germany\\
$^5$Key Laboratory for Computational Astrophysics, National Astronomical Observatories, Chinese Academy of Sciences, Beijing 100101, China\\
}

\date{Accepted XXX. Received YYY; in original form ZZZ}

\pubyear{2025}

\begin{document}
\label{firstpage}
\pagerange{\pageref{firstpage}--\pageref{lastpage}}
\maketitle

\begin{abstract}
 We present a cosmological zoom-in simulation targeting the high redshift compact progenitor phase 
 of massive galaxies, with the most massive galaxy reaching a stellar mass of $M_{\star}=8.5\times 10^{10} \ M_{\odot}$ at $z=5$. The dynamics of supermassive black holes (SMBHs) is modelled from seeding down to their coalescence at sub-parsec scales due to gravitational wave (GW) emission by utilising a new version of the \ketju{} code, which combines regularised integration of sufficiently massive SMBHs with a dynamical friction subgrid model for lower-mass SMBHs. All nine massive galaxies included in this study go 
 through a gas-dominated phase of early compaction in the redshift range of $z\sim 7-9$, starting at stellar masses of $M_\star\gtrsim 10^8\ \mathrm{M}_\odot$ and ending at a few times $M_{\star}\sim 10^9\ \mathrm{M}_\odot$. The sizes, masses and broad band fluxes of these compact systems are in general agreement with the population of systems observed with JWST known as `Little Red Dots'. In the compact phase, the stellar and SMBH masses grow rapidly, leading to a sharp decline in the central gas fractions. The outer regions, however, remain relatively gas-rich, leading to subsequent off-centre star formation and size growth. Due to the very high 
 central stellar densities ($\rho_{\star}\gtrsim 10^{13}\,\mathrm{M_\odot/kpc^3}$), the SMBHs merge rapidly, typically just $\sim 4-35\ \mathrm{Myr}$ after the SMBH binaries have become bound. Combining \ketju{} with the phenomenological PhenomD model resolves the complete evolution of the GW emission from SMBH binaries through the Pulsar Timing Array frequency waveband up to the final few orbits that produce GWs observable with the future LISA mission. 
\end{abstract}




\section{Introduction}

Supermassive black holes (SMBHs) with masses in the range of $m_{\bullet}=10^{6}-10^{10} \ M_{\odot}$ are found at the centres of most if not all massive galaxies in the Local Universe (see e.g. \citealt{Kormendy2013}). In the past few decades a number of observational programs, including, but not limited to the Sloan Digital Sky Survey, PANSTARRS1 and the Wide-field Infrared Survey (WISE) have also discovered a population of bright quasars (bolometric luminosities of $L_{\rm bol}\sim 10^{47} \ \rm erg \ s^{-1}$) at moderately high redshifts of $z\sim 4-7$ (e.g. \citealt{2001AJ....122.2833F,2011Natur.474..616M,2016ApJS..227...11B,2017MNRAS.468.4702R,2018ApJ...869..150M,2019AJ....157..236Y}).  

Recently, the James Webb Space Telescope (JWST) has revolutionised the studies of the high-redshift Universe by discovering in addition to the bright quasars a population of lower luminosity Active Galactic Nuclei (AGNs) 
at redshifts of $z\sim 4-7$ (e.g. \citealt{2023ApJ...942L..17O,2023A&A...677A.145U,2023ApJ...954L...4K,2024A&A...691A.145M,2024ApJ...964...39G,2025ApJ...991...37A}). Some of these objects belong to a new class of objects unseen before, which have been called 
''little red dots'' (LRDs), due to their pointlike morphology, red colours in the observed frame $\sim 2-5 \ \mu \rm m$, and often flat or even blue colours from $1-2 \ \mu \rm m$ (e.g. \citealt{Akins2023,Akins25,Baggen23,Baggen2024,2024ApJ...963..129M,Kokorev24,Ma2025_lrd}). 
In addition, JWST has also observed a handful, but steadily increasing number of AGNs at very high redshifts of 
$z\sim 9-11$ (e.g. \citealt{2023ApJ...953L..29L,2024A&A...691A.145M,2024NatAs...8..126B,2025ApJ...989L...7T}). 
Finally, JWST has also revealed that the rate of dual AGNs at high redshifts is significantly higher than the predictions made from cosmological simulations \citep{2025A&A...696A..59P,2024MNRAS.531..355U,2025arXiv250921575U}. 

The origins of the SMBHs powering the AGNs and how SMBHs are able to grow to such large masses on relatively short timescales are also still uncertain (e.g. \citealt{2024OJAp....7E..72R}). It is possible that the SMBHs originate from the remnants of population-III stars (so-called light seeds; \citealt{2001ApJ...551L..27M} ), which would have large number densities. Another possibility is that SMBHs form directly from the collapse of massive gas clouds (forming so-called heavy seeds). Such a formation channel requires atomic cooling halos, which would need Lyman--Werner radiation and low metallicities in order to constrain the formation of molecular hydrogen (e.g. \citealt{2020ARA&A..58...27I}). This would also result in a smaller abundance of SMBHs compared to the light seed scenario \citep{2024OJAp....7E..72R}. 

In the $\Lambda$CDM hierarchical model galaxies grow through galaxy mergers and gas accretion, with the mergers of SMBHs proceeding generally through a three-stage process \citep{1980Natur.287..307B}. Firstly, at large kpc scales, the unbound SMBH pair sinks toward the centre of the merger remnant through dynamical friction \citep{Chandrasekhar43}. The SMBH binary will eventually first become bound and then later 
transition into a `hard' binary, when the binding energy of the binary exceeds the average kinetic energy of the surrounding stars \citep{2013degn.book.....M}. The binary separation is typically $a\sim 1-10 \ \rm pc$ at this second stage and the further shrinking of the binary orbit will then primarily proceed by the scattering of individual stars, residing in the so called `loss-cone' \citep{Quinlan96,Sesana2006}. At this stage, in gas-rich environments the additional drag from gas in the form of a circumbinary disc can also impact the orbital evolution of the binary (e.g. \citealt{2014ApJ...783..134F,Duffel_formula}). Finally, in the third and final stage, the SMBH binary will be driven to coalescence by gravitational wave (GW) emission, typically at sub-parsec separations \citep{PetersMathews}. 

GWs are now routinely observed from the mergers of stellar-mass $(m_{\bullet}\lesssim 100 \ \mathrm{M}_{\odot})$ black holes (BHs) by the LIGO-Virgo-KAGRA (LVK) collaboration \citep{2023PhRvX..13d1039A}. Recent detections by the LVK collaboration are also starting to probe the regime of intermediate-mass BHs, with a 
detection of a binary merger with a final BH mass in excess of $m_{\bullet}\gtrsim 225 \ \mathrm{M}_{\odot}$ \citep{2025ApJ...993L..25A}. GWs from more massive BHs have not as of yet been directly detected. However, observations by pulsar timing arrays (PTAs) have already provided tentative evidence for a stochastic GW background in the nano-Hertz frequency regime, which is likely sourced by a population of very massive $(m_{\bullet}\gtrsim 10^{8}-10^{9} \ \mathrm{M}_{\odot})$ merging SMBHs in the relatively local ($z\lesssim 1$) Universe \citep{Nanograv15_obs_and_timing,2023A&A...678A..50E,2023RAA....23g5024X,2023PASA...40...49Z}. The next decade will also see the launch of the Laser Interferometer Space Antenna (LISA), which will be capable of detecting directly the GW signal from the mergers of SMBHs in the mass range $m_{\bullet}\sim 10^{4}-10^{7} \ \mathrm{M}_{\odot}$ potentially up to redshifts as high as $z\sim 20$ \citep{2023LRR....26....2A}.  

In numerical simulations the dynamical evolution of SMBHs is commonly modelled using a repositioning technique, in which the SMBH is repositioned on the local minimum of the potential at every timestep \citep{Volker_thermal,Johansson2009,2022MNRAS.516..167B}. Rather than modelling the orbital evolution of the SMBH, the aim of this method is to keep the SMBHs near the centre of the galaxy, with this method typically being employed in simulations where the SMBH and baryonic particles have similar masses (e.g. \citealt{2009ApJ...707L.184J,2015MNRAS.446..521S,2017MNRAS.465.3291W}). Such a treatment of SMBHs was necessary as SMBHs would have otherwise been in off-centre regions with low gas densities, and would thus be unable to efficiently quench star formation via feedback. Repositioning methods also tend to sink SMBHs too rapidly in haloes and in addition result in artificially-enhanced merging of SMBHs, affecting particularly low-mass seed BHs at high redshifts \citep{Ma_seeds_2021,Chen_DF_model, 2025MNRAS.542.2019B}. 

Another fundamental limitation of hydrodynamical galaxy formation simulations is the necessary inclusion of gravitational softening or equivalently a minimum grid cell that sets the minimum spatial resolution, below which the dynamics of SMBHs cannot be resolved. A commonly adopted approach that circumvents the effects of softening in the dynamical friction phase is to add a subresolution drag force to the equations of motion that accounts for the unresolved dynamical friction (e.g. \citealt{Tremmel_subgrid_2015,2019MNRAS.486..101P,Chen_DF_model,Ma_discrete_subgrid_2023,Genina_DF,Damiano2025}). Adding a subgrid dynamical friction force also removes the need for repositioning, thus enabling a more accurate representation of the large-scale SMBH dynamics. However, in this approach the SMBHs are still numerically merged at an unphysically large separation, typically corresponding to a gravitational softening length (e.g. \citealt{2011ApJ...742...13B,2025MNRAS.542.2019B,2025ApJ...991...81B,2025arXiv251001322B}). In addition, subgrid models require the SMBH mass to be larger than the surrounding baryonic and dark matter particles, thus limiting the mass of the seeded SMBHs.

Cosmological hydrodynamical simulations have been essential in modelling the formation and merging history of SMBHs throughout cosmic history (e.g. \citealt{2015MNRAS.452..575S,Habouzit2021}). In these simulations the SMBHs are typically seeded at some fixed mass, once the dark matter halo mass has reached a critical mass \citep{2007MNRAS.380..877S,2008ApJ...676...33D}. However, there are also more sophisticated seeding mechanisms that employ additional constraints such as a sufficiently low metallicity threshold and a critical Lyman Werner flux, as to promote the formation of massive direct collapse BHs, with some models also adopting a distribution of seed BH masses, instead of a fixed mass (e.g. \citealt{2014ApJ...795..137R,2020MNRAS.492.3021R,astrid_smbhs,2024MNRAS.531.4311B}). 

The fundamental limitation of gravitational softening in resolving the small-scale dynamics of SMBHs is well recognised. This shortcoming is typically addressed by applying semi-analytic post-processing techniques to estimate the time delays at sub-kpc spatial scales based on the resolved global properties of the host galaxies (e.g. \citealt{2017MNRAS.464.3131K,Kelley2017,2019MNRAS.486.4044B,2021MNRAS.501.2531S,2022ApJ...933..104L,2022MNRAS.509.3488I}).   
Various approaches have been used to further model SMBH binaries from softened simulations to GW-driven regime. A common method is to re-simulate a selected volume from cosmological simulation (usually containing one galaxy merger) using dedicated direct $N$-body codes \citep{Khan2016,Chen2024_magicsI,Fastidio2024,2025ApJ...981..203M}. It is also possible to resimulate a cosmological volume with a regularised integrator \citep{Mannerkoski2021, Mannerkoski2022} or to use such a model for a resimulated merger with higher mass resolution \citep{2025ApJ...980...79Z}. Another recent approach is to (similarly to the resimulations with a regularised integrator) use softened gravity for large distance interactions, but model SMBH binaries via a subgrid analytic model \citep{Ramcoal}.

The \ketju{} code \citep{2017ApJ...840...53R,Gadget4-ketju} is an example of a hybrid approach in which the softened \textsc{gadget-3} \citep{2005MNRAS.364.1105S} code was initially combined with a parallelised re-implementation of the regularised \textsc{ar-chain} \citep{2008AJ....135.2398M} integrator. The code was initially used to run isolated galaxy mergers \citep{rantala2018,2019ApJ...872L..17R,Rantala_nuggets,2023MNRAS.526.2688R,2025MNRAS.537.3421R,2025ApJ...991...83R}. The \ketju{} code was subsequently improved by replacing the \textsc{ar-chain} integrator with the \textsc{mstar} integrator \citep{MSTAR}, which resulted in a significant performance improvement thanks to a novel efficient two-fold parallelization scheme, thus enabling also hydrodynamical simulations with accurate small-scale dynamics. The \ketju{} code has since been used to study the dynamics of SMBHs in both cosmological zoom-in simulations \citep{Mannerkoski2021,Mannerkoski2022} and in high-resolution hydrodynamical simulations, including a model for SMBH binary accretion and feedback \citep{Liao_binary_feedback_2023,Liao_rabbits_I_2024,Liao_rabbits_II_2024}. In addition, the code has been used to study the nuclear dynamics of intermediate mass black hole growth as well as the 
formation and evolution of stellar clusters in high-resolution dwarf galaxy simulations that resolve individual stars \citep{Partmann_mergers,2025MNRAS.537..956P,2025MNRAS.538.2129L,2025MNRAS.543.1023L}.  

\ketju{} simulations that employ macroparticles, with typical masses of $m_{\star}\sim 10^{5}$ require 
a sufficiently high BH to stellar particle mass ratio $(m_{\bullet}/m_{\star}\gtrsim 100)$, in order to accurately resolve the BH-stellar particle interactions. Because of this requirement, previous cosmological simulations run with \ketju{} presented in \citet{Mannerkoski2021,Mannerkoski2022} could only use the accurate \ketju{} integration at relatively low redshifts of $z\lesssim 1$, when the SMBHs had grown to be sufficiently massive. 
Prior to turning on \ketju{}, these simulations were run using standard \textsc{gadget-3}, including BH repositioning. In this paper, on the contrary, we develop and test a new version of the \ketju{} code, in which the code is combined with a subgrid dynamical friction model. The combination of \ketju{} with a subgrid dynamical friction model allows us to run simulations without repositioning and with \ketju{} integrated dynamics for massive BHs from the start of cosmological simulations at high redshifts, while simultaneously following global galactic-scale 
dynamical and astrophysical processes in the high-redshift compact gas-rich galaxies, hosting these SMBHs.

This paper is structured as follows. In Section \ref{sect:Num_methods}, we first briefly review the main features of the \ketju{} code and then proceed to discuss how the dynamical friction subgrid model is implemented in the code. In Section \ref{sect:plummer_sphere} we then perform dynamical friction runs in a Plummer sphere, testing the efficiency of SMBH sinking in different dynamical friction models, comparing the results also to \ketju{}. Next, in Section \ref{sect:brownian} we study the effect of Brownian motion in isolated simulations set in a Hernquist sphere. The test simulations of SMBH sinking and Brownian motion are necessary for the application of the \ketju{} code combined with a dynamical friction subgrid model in a cosmological setting. We then perform cosmological zoom-in simulations using the new combined model and study in Section \ref{sect:gal_evo} the properties of the host galaxies of the massive SMBHs. Here we also compare our simulations with observed galaxies. This is followed by Section \ref{sect:cosmo_bh} in which we study the dynamics and mergers of the SMBH binaries hosted by the galaxies. In Section \ref{sect:discussion} we discuss our results and in Section \ref{sect:conclusions} we finally present our conclusions.

\section{Numerical methods}
\label{sect:Num_methods}
\subsection{The KETJU code}

The \ketju{} code\footnote{The public version of the \ketju{} code based on \textsc{gadget-4} can be accessed at \href{https://www.mv.helsinki.fi/phjohans/ketju}{https://www.mv.helsinki.fi/phjohans/ketju}} \citep{2017ApJ...840...53R,Gadget4-ketju} extends the widely used  \textsc{gadget-3}/\textsc{gadget-4} codes \citep{2005MNRAS.364.1105S,2021MNRAS.506.2871S} by replacing the standard leapfrog integration with the algorithmically regularised higher accuracy \textsc{mstar} integrator \citep{MSTAR} that is used to solve the dynamics in small
regions in the vicinity of BHs. The size of the regularised \ketju{} region is set at three times the BH gravitational softening length $(r_\mathrm{Ketju}=3\times \varepsilon_{\bullet})$ in order to ensure that all BH-BH and all BH-star interactions are always non-softened. 

In order to account for relativistic effects, such as the emission of GWs the \ketju{} code includes post-Newtonian (PN) corrections for SMBH binaries up to and including order 3.5PN \citep{2004PhRvD..69j4021M,PN_terms}. In addition 
to binary PN terms, we also include the 
leading order 1PN corrections of general $N$-body systems, which contain terms involving up to three SMBHs that could potentially affect the long-term evolution of triplet SMBH systems (\citealt{2014PhRvD..89d4043W}, see also \citealt{Mannerkoski2022} for further details). 

\subsection{Hydrodynamics and stellar physics}

The gas hydrodynamics and stellar physics is modelled using the same models as in \citet{Mannerkoski2021,Mannerkoski2022}. For gas we use the SPHGal smooth particle hydrodynamics (SPH)   
implementation \citep{Hu2014}, which employs a pressure-entropy formulation, including artificial viscosity and artificial conduction, and utilises a  $C^4$ Wendland kernel with $N_\mathrm{ngb}=100$ neighbours.

The adopted model for gas cooling and stellar physics in the \ketju{} code follows the descriptions initially 
developed in \citet{Scannapieco05, Scannapieco06} and later refined in \citet{Aumer,2017ApJ...836..204N}. In this model the abundances of 11 chemical species are tracked for every gas and stellar particle. The cooling rate of 
each gas particle depends on its temperature, density and chemical composition. The cooling rates are adopted from tables 
in \cite{Wiersma}, which assume that the gas is optically thin, in ionisation equilibrium and embedded in 
a redshift dependent UV/X-ray background from quasars and galaxies \citep{2001cghr.confE..64H}. 

Stars are formed stochastically from gas particles that fulfill the star formation criteria: a hydrogen number density above the critical density of $n_{\rm H}\ge 0.1 \ \rm cm^{-3}$, a temperature below $T \le 12 000 \ \rm K$, with the gas residing in a convergent gas flow, i.e., $\nabla \cdot \mathbf{v}_{\rm gas} \le 0$. The probability then to convert a gas particle into a stellar particle is given by   
\begin{equation}
\begin{split}
    p_\mathrm{SF} & =1-\exp\left(-\epsilon_\mathrm{SFR}\frac{\Delta t}{t_{\rm dyn}} \right) \\
    & = 1 - \exp\left( -\epsilon_\mathrm{SFR}\Delta t \sqrt{4\pi G \rho_\mathrm{gas}} \right),
\end{split}
\end{equation}
where $\rho_\mathrm{gas}$ is the gas density, $\Delta t$ the timestep, $G$ the gravitational constant and $\epsilon_\mathrm{SFR}=0.02$ is the star formation rate efficiency. 

Finally, the model also includes stellar feedback from the explosions of Type II (SNII) and Type I (SNIa) supernovae and the slow winds from asymptotic giant branch (AGB) stars. Stellar particles flagged for feedback deposit both thermal and kinetic feedback to the closest 10 neighbouring particles, while simultaneously enriching the gas with metals, with the SNII, SNIa and AGB yields derived from \citet{1995ApJS..101..181W,1999ApJS..125..439I,2010MNRAS.403.1413K}, respectively. 

Each stellar particle represents a stellar population with a \citet{1955ApJ...121..161S} initial mass function. We assume that all stars more massive than 8 $M_{\odot}$ end their lives in SNII events and we approximate that all SNII in a star particle explode at the same age of $\tau_{\rm SN II}=3 \ \rm Myr$ in one-off events. For SNIa, in contrast, which represent carbon oxygen white binary systems, we assume that the first SNIa in a star particle explodes at the age of $\tau_{\rm SN Ia, min}=50 \ \rm Myr$, with the following SNIa events performed at every 50 Myr, until a maximum time period of $\tau_{\rm SN Ia, max}=10 \ \rm Gyr$ is reached. When a star undergoes a feedback event, the mass $m_{\rm ej}$ is ejected with an outflow velocity of $v_{\rm ej}$, resulting in a total energy of 
\begin{equation}
    E_{\rm SN}=\frac{1}{2}m_{\rm ej} v_{\rm ej}^{2}, 
\end{equation}
with the ejection velocity set at $v_{\rm ej}=4000 \ \rm km/s$ for type II and Ia supernovae, and at the much lower value of $v_{\rm ej}=25 \ \rm km/s$ for the winds of AGB stars. For further details on the implementation of gas cooling, star formation and stellar feedback see \citet{Liao_binary_feedback_2023}. 

\subsection{SMBH accretion and feedback}

For SMBHs, we use the prescription introduced in \cite{Liao_binary_feedback_2023} (see also \citealt{Liao_rabbits_I_2024, Liao_rabbits_II_2024}). Two SMBHs are bound when their total orbital energy is negative, i.e.
\begin{equation}
    E=\frac{1}{2}\mu v^2-G\frac{m_1m_2}{r}<0,
\end{equation}
where $m_{1,2}$ are the masses of the two SMBHs, $v$ and $r$ are their relative velocity and separation, respectively, and $\mu=m_1m_2/(m_1+m_2)$ is the reduced mass.

When a SMBH is not in a bound binary, the accretion follows the standard Bondi--Hoyle--Lyttleton (BHL) accretion model \citep{Bondi44, Bondi52, Hoyle-Lyttleton}. The maximum accretion rate is given by the Eddington limit, 
\begin{equation}
    \dot{m}_\mathrm{Edd}=\frac{4\pi Gm_\bullet m_\mathrm{p}}{\epsilon_\mathrm{r}\sigma_\mathrm{T}c}, 
\end{equation}
where $m_\mathrm{p}$ is the mass of a proton, $\sigma_\mathrm{T}$ is the Thomson cross-section, $c$ is the speed of light and $\epsilon_\mathrm{r}$ is the radiative efficiency, set to 0.1 in this study \citep{Liao_binary_feedback_2023}. This requirement of the Eddington limited rate results in an accretion rate of $\dot{m}_\bullet=\mathrm{min}(\dot{m}_\mathrm{Edd},\dot{m}_\mathrm{BHL})$ with
\begin{equation}
\dot{m}_\mathrm{BHL}= \alpha\frac{4\pi G^2m_\bullet^2\rho}{\left(c_\mathrm{s}^2+v_\mathrm{rel}^2\right)^{3/2}},
\end{equation}
where $\rho$ and $c_\mathrm{s}$ are the density and the sound speed of the gas surrounding the SMBH, respectively, $v_\mathrm{rel}$ is the SMBH velocity relative to the gas and $\alpha$ is a dimensionless boost factor. This factor is necessary due to the limited numerical resolution and is set to $\alpha=25$ \citep{2009ApJ...707L.184J}. The gas properties are calculated within a SPH smoothing length $h_{\rm SPH}$. 

When two SMBHs are in a bound binary, we instead adopt the \cite{Liao_binary_feedback_2023} binary model, in which 
the single SMBH is replaced by the properties of the binary in the accretion formulae via a circumbinary disc subgrid model. The disc is centred at the centre-of-mass (CoM) of the binary. The total accretion rate of the binary is limited by the Eddington rate
\begin{equation}
    \dot{m}_\mathrm{Edd,bin}=\frac{4\pi Gm_\mathrm{bin}m_\mathrm{p}}{\epsilon_\mathrm{r}\sigma_\mathrm{T}c},
\end{equation}
where $m_\mathrm{bin}=m_1+m_2$ is the total mass of the two BHs. The BHL formula then becomes
\begin{equation}
    \dot{m}_\mathrm{BHL,bin} = \alpha\frac{4\pi G^2m_\mathrm{bin}^2\rho_\mathrm{CoM}}{\left(c_\mathrm{s,CoM}^2+v_\mathrm{rel,CoM}^2\right)^{3/2}},
\end{equation}
where and $v_\mathrm{rel,CoM}$ is the CoM velocity relative to the gas, $\rho_\mathrm{CoM}$ and $c_\mathrm{s,CoM}$ are the density and sound speed at the CoM of the binary, respectively. Adopting the gas properties at the CoM here is essential, as the high inspiral SMBH velocities in the tight binary phase would otherwise artificially suppress accretion rates and AGN feedback, leading to unphysical evolutionary jumps in these quantities (see \citealt{Liao_binary_feedback_2023} for details). The total mass accretion of the binary, $\dot{m}_\mathrm{bin}=\dot{m}_1+\dot{m}_2$ is then divided between the two SMBHs using the formula given by \citet{Duffel_formula},
\begin{equation}
    \frac{\dot{m}_2}{\dot{m}_1}=\frac{1}{0.1+0.9q},
\label{eq:duffell}
\end{equation}
where $q=m_2/m_1\leq1$. Importantly, in this model the smaller, secondary SMBH of a binary always has a higher accretion rate than the main SMBH.

The SMBH feedback is modelled as thermal feedback (see e.g. \citealt{DiMatteo_thermal, Volker_thermal}). A fraction $\epsilon_\mathrm{f,th}$ of the radiated energy is coupled to the surrounding gas within the SPH smoothing length $h_{\rm SPH}$. The radiated luminosity is 
\begin{equation}
    \dot{E}_\mathrm{th}= \epsilon_\mathrm{f,th}\epsilon_\mathrm{r}\dot{m}_\bullet c^2
\end{equation}
with a gas particle $i$ receiving an amount of
\begin{equation}
    E_{\mathrm{th,}i}=\frac{m_iw_i\dot{E}_\mathrm{th}\Delta t}{\rho},
\end{equation}
where $w_i$ is SPH kernel weight of the particle, $m_i$ is the mass of the particle and $\Delta t $ the timestep. Following \citet{Liao_binary_feedback_2023}, a value of $\epsilon_\mathrm{f,th}=0.02$ is chosen for the feedback efficiency.

\subsection{Dynamical friction subgrid models}
\label{sect:dyn_fric_subgrid}

A SMBH moving through a field of less massive particles (e.g. stars) pulls nearby particles towards itself, generating an overdense wake behind the trajectory of the massive particle. The overdense region then deaccelerates the massive particle, an effect known as dynamical friction \citep{Chandrasekhar43}.

Modelling dynamical friction in cosmological simulations is challenging, in particular at high redshifts when the particles representing the recently seeded SMBHs are often only a few times the mass of the background particles. While many modern simulations use repositioning, e.g. IllustrisTNG \citep{2018MNRAS.473.4077P}, FLAMINGO \citep{2023MNRAS.526.4978S} and COLIBRE \citep{2025arXiv250821126S}, some simulations, e.g. ROMULUS \citep{Romulus}, NEWHORIZON \citep{Newhorizon} and ASTRID \citep{astrid_smbhs} resolve the issue by applying a subgrid correction to the acceleration of SMBHs in order to account for the unresolved dynamical friction. Here, we briefly review the main aspects of two such models, namely the 
\cite{Tremmel_subgrid_2015} and \cite{Ma_discrete_subgrid_2023} models before proceeding to discuss how such
subgrid dynamical friction models can be combined with the \ketju{} code. 

In the \cite{Tremmel_subgrid_2015} model under the assumption that the velocity distribution of particles around the SMBH is isotropic within the gravitational softening length $\varepsilon_\bullet$, a SMBH with mass $m_\bullet$ gains an additional acceleration due to dynamical friction given by the \cite{Chandrasekhar43} formula
\begin{equation}
\label{eq:a_df_tremmel}
    \mathbf{a}_\mathrm{df} = -4\pi G^2m_\bullet\rho(<v_\bullet)\ln\Lambda\frac{\mathbf{v}_\bullet}{v_\bullet^3}.
\end{equation}
Here $\ln\Lambda$ is the Coulomb logarithm, $\rho(<v_\bullet)$ is the density of collisionless (dark matter and stellar) particles moving slower than the SMBH relative to the velocity of the local centre-of-mass $v_\bullet$ ($v_\bullet=|\mathbf{v}_\bullet|$). Since the velocity distribution is isotropic, this density can be calculated as
\begin{equation}
    \rho(<v_\bullet)=\frac{M(<v_\bullet)}{M}\rho,
\end{equation}
with the density and the masses calculated using the hundred nearest collisionless particles. The Coulomb logarithm is evaluated using the minimum and maximum impact parameters, i.e.
\begin{equation}
    \ln\Lambda=\ln \left(\frac{b_\mathrm{max}}{b_\mathrm{min}}\right)=\ln \left(\frac{\varepsilon_\bullet}{\mathrm{max}(b_{90},R_\mathrm{Sch})}\right),
\end{equation}
where the maximum impact parameter $b_\mathrm{max}=\varepsilon_\bullet$ is set to the SMBH softening length. The minimum impact parameter is set to $b_\mathrm{min}=\mathrm{max}(b_{90},R_\mathrm{Sch})$, where $R_\mathrm{Sch}$ is the Schwarzschild radius 
and $b_{90}$ is the 90 degree deflection radius defined as
\begin{equation}
    b_{90}=\frac{Gm_\bullet}{v_\bullet^2}.
\end{equation}

An alternative model for dynamical friction that, unlike most subgrid models based on the \citet{Chandrasekhar43} model, does not assume an isotropic and homogeneous medium of field particles was recently developed by \cite{Ma_discrete_subgrid_2023}. The authors instead present a discrete dynamical friction formula, that allows the full dynamical friction force from all particles to be computed self-consistently by direct summation. With a mass ratio of 10 between a SMBH and the surrounding particles, they show that the model results in SMBH trajectories and velocities matching the estimated velocity acquired with \citet{Chandrasekhar43}. The total dynamical friction acceleration of an SMBH due to other particles is given by

\begin{equation}
    \mathbf{a}_{\mathrm{df},\bullet}=\sum_i \mathbf{a}_{\mathrm{df},i \rightarrow \bullet}=\sum_i\frac{\alpha_ib_i}{(1+\alpha_i^2)r_i}\left(S_i(r_i) \frac{Gm_i}{r_i^2}  \right)\hat{\mathbf{v}}_i,
\end{equation}
where $m_i$ is the mass of particle $i$, $r_i$ and $\hat{\mathbf{v}}_i=\frac{\mathbf{v}_i}{v_i}$ are the separation of the particle from the SMBH and the unit relative velocity, respectively. $S_i(r)$ is the gravitational softening kernel of particle $i$ at the separation $r_i$. The summation is done over all particles encountered in the gravitational force calculation. Finally, assuming the limit $m_{\bullet}\gg m_i$, i.e. that the BH is much more massive than the background field particles causing the dynamical friction (at least by a factor of a few), the $\alpha_{i}$-parameter can be expressed as,  

\begin{equation}
\alpha_i \approx b_i \frac{v_i^2}{Gm_\bullet}
\end{equation}
and
\begin{equation}
    b_i = r_i|\hat{\mathbf{r}}_i- (\hat{\mathbf{r}}_i\cdot \hat{\mathbf{v}}_i)\hat{\mathbf{v}}_i|,
\end{equation}
where $\hat{\mathbf{r}}_i=\frac{\mathbf{r}_i}{r_i}$ and $b_{i}$ is the impact parameter of the interaction. 

In contrast to other dynamical friction subgrid models based on the integrated formula of \citet{Chandrasekhar43}, the direct summation method enables the manifest conservation of momentum. This is done by applying an acceleration term to particles other than SMBHs according to
\begin{equation}
    \mathbf{a}_{\mathrm{df},\bullet\rightarrow i} = -\frac{m_\bullet}{m_i}\mathbf{a}_{\mathrm{df},i \rightarrow \bullet}.
\end{equation}
One drawback of this approach is that in relatively well-resolved cases the model effectively double counts the dynamical friction because a fraction of the force is already resolved in the simulation and is thus already accounted for in the regular force calculation. This shortcoming is acknowledged by \cite{Ma_discrete_subgrid_2023}, who point towards some mass-resolution dependent correction functions that could be implemented (see also the discussion in \citealt{Genina_DF}).

\subsection{Combining \textsc{ketju} with a dynamical friction subgrid model}
\label{sect:K+T}

\begin{figure}
    \centering
    \includegraphics[width=1\linewidth]{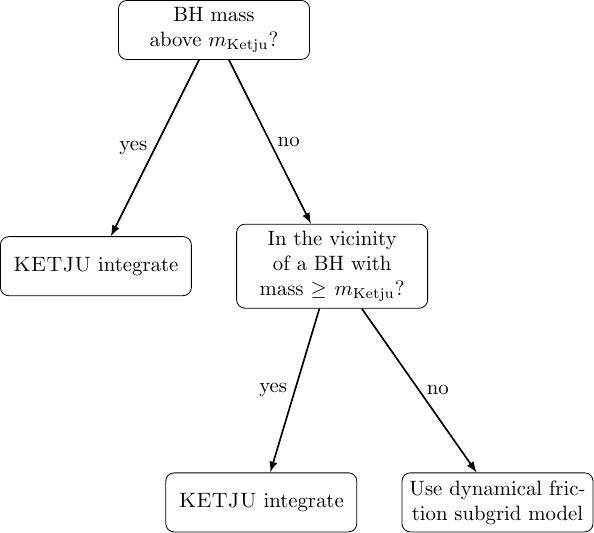}
    \caption{A dichotomous key describing whether \ketju{} or the dynamical friction subgrid model is used for the BH dynamics. The dynamical friction subgrid model is used when a SMBH has a mass of $m_\bullet<m_\mathrm{Ketju}$ and the separation from a \ketju{} integrated SMBH is larger than $10\times r_\mathrm{Ketju}$.}
    \label{fig:flowchart}
\end{figure}

In a typical galaxy formation simulation which covers a cosmic time of billions of years the stellar component is modelled using macroparticles with typical stellar masses of $m_{\star}\sim 10^{4}-10^{6} \ M_{\odot}$. The \ketju{} code is able to resolve accurately the dynamical interactions between SMBHs and the stellar component, but this requires that the mass ratio between the SMBH and stellar particles is sufficiently high, for which cosmological simulations have used a mass ratio of the order of $m_\bullet/m_\star\sim 100$ \citep{Mannerkoski2022}. However, we note that an even higher mass ratio of $m_\bullet/m_\star\gtrsim 1000$ would be typically required for a fully converged binary hardening rate \citep{Gadget4-ketju}, although \citet{2017ApJ...840...53R} show that the evolution of the inverse binary semimajor axis does not differ in  galaxy mergers that include both stellar and dark matter components for particle mass ratios in the range $m_\bullet/m_\star\sim 200-2000$. 

For too low particle mass ratios the individual interactions between SMBH and the stellar macroparticles would be unphysically strong leading to inaccurate dynamics, as shown in sections \ref{sect:plummer_sphere} and \ref{sect:Brownian}. In order to model accurately the large-scale dynamics of low-mass SMBHs, for which $m_\bullet/m_\star\lesssim 100$ the \ketju{} code needs to 
be combined with a dynamical friction subgrid model. In this combined hybrid model an analytic subgrid dynamical friction model is used for low-mass SMBHs, which will eventually transition to become \ketju{} integrated SMBHs once their mass has grown above the \ketju{} mass limit.

In Figure \ref{fig:flowchart} we show a dichotomous key describing the combined \ketju{} dynamical friction model. During each timestep when integrating a SMBH, we first check its mass and compare it to the preset \ketju{} mass limit $m_\mathrm{Ketju}$. If the mass of the SMBH is above the \ketju{} mass limit the SMBH and the surrounding stellar component (within a radius of $r_\mathrm{Ketju}=3\times\varepsilon_\bullet$) will be integrated with the \textsc{ketju} integrator. In addition, the \textsc{ketju} integrator will also be used if the SMBH mass is below the mass limit, but the SMBH is within a distance of $10\times r_\mathrm{Ketju}$ from a \ketju{} BH, with mass $m_\bullet \geq m_\mathrm{Ketju}$. If neither of these two criteria are met, the dynamical friction subgrid model of \citet{Tremmel_subgrid_2015} will be used instead.

\ketju{} integrated SMBH binaries are numerically merged when their separation is twelve times the 
Schwarzschild radius, corresponding to the total binary mass \citep{Gadget4-ketju}. For SMBHs that are below the 
\ketju{} mass limit and are integrated using the standard \textsc{gadget-3} integrator including the dynamical friction model, we use the standard \textsc{gadget-3} merger criterion, in which the two SMBHs are numerically merged, when they are within the SPH smoothing length $h_{\rm SPH}$ from each other, and their relative velocity is smaller than half of the local sound speed \citep{2005MNRAS.364.1105S,Johansson2009}. The standard \textsc{gadget-3} merger criterion results in numerical mergers at separations of $\sim 100\ \mathrm{pc}$, which is typically $\sim 5-6$ orders of magnitude larger than the \ketju{} merger separation.

\begin{figure*}
    \centering
    \includegraphics[width=\linewidth]{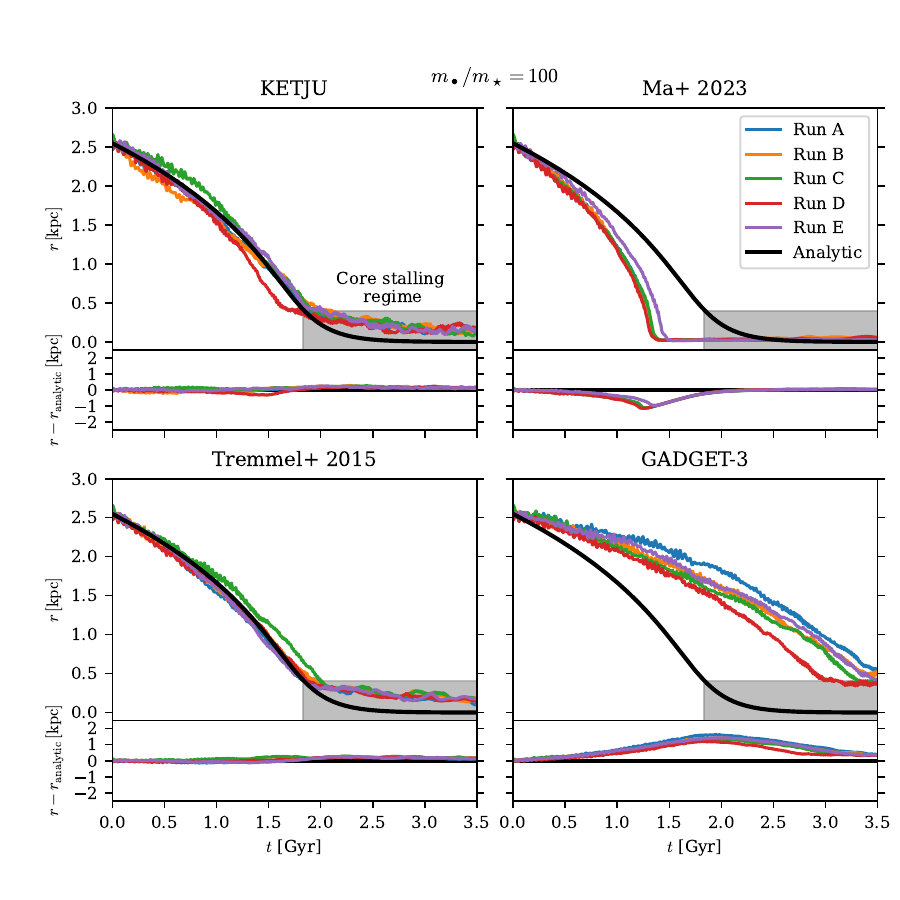}
    \caption{The sinking of a SMBH into the centre of a Plummer sphere. The mass ratio between the SMBH and a stellar particle is $m_\bullet/m_\star=100$ in each simulation. For each SMBH dynamics model, the separation from the centre (main panels) and the residual (smaller panels) from the analytical estimate is shown. The lines are averaged over $25\ \mathrm{Myr}$. A shaded gray area is the region where \ketju{} begins to deviate from the analytical estimate due to core stalling not captured with the analytical model. The analytical estimate is best matched by \ketju{} and the subgrid model from \citet{Tremmel_subgrid_2015}.}
    \label{fig:plummer_100}
\end{figure*}

\begin{figure*}
    \centering
    \includegraphics[width=\linewidth]{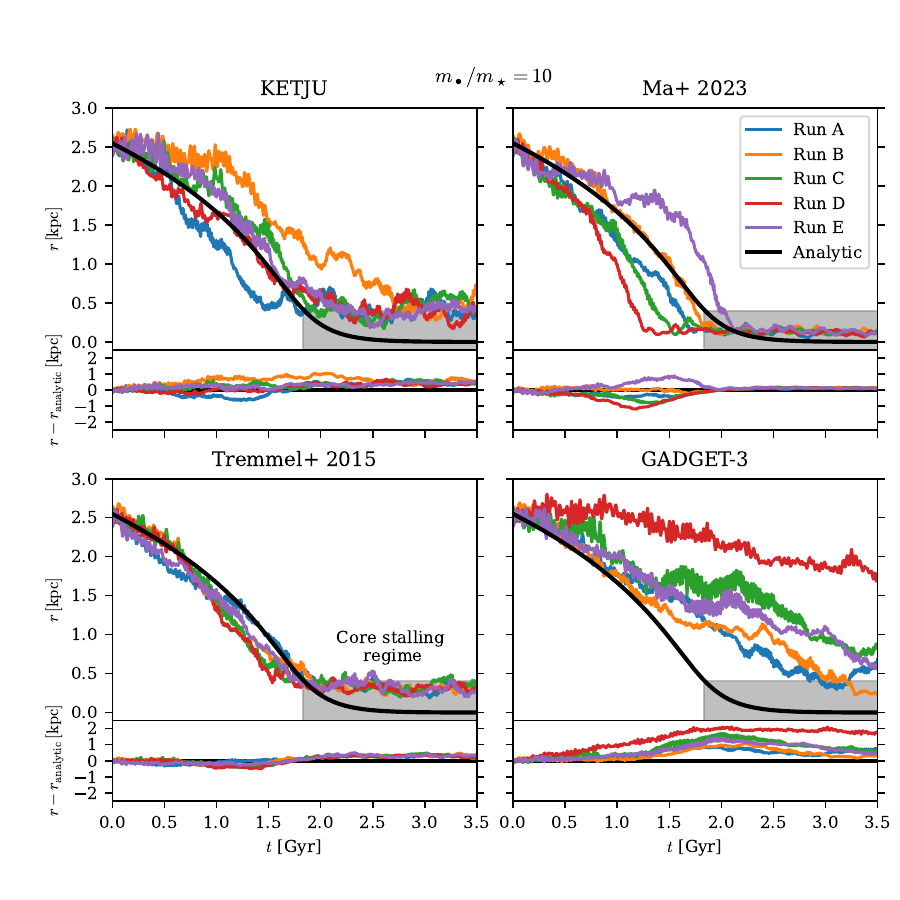}
    \caption{The sinking of a SMBH into the centre of a Plummer sphere, i.e. the same as Fig. \ref{fig:plummer_100}, but now with a mass ratio of $m_\bullet/m_\star=10$. The analytic estimate is best matched by the subgrid model of \citet{Tremmel_subgrid_2015}. 
    }
    \label{fig:plummer_10}
\end{figure*}

\section{Dynamical friction in a Plummer sphere}
\label{sect:plummer_sphere}

In order to apply the combined version of \ketju{} and a dynamical friction subgrid model in a cosmological setting, we first need to understand the required mass ratio at which \ketju{} is capable of resolving the dynamics of single SMBHs. Therefore we first run low resolution test simulations of isolated galaxies including one SMBH and only collisionless particles. Two sets of test simulations are produced, focusing on SMBH sinking
(Section \ref{sect:plummer_sphere}) and Brownian motion (Section \ref{sect:brownian}). 

A specific model for SMBH dynamics can only be used when the mass ratio between a SMBH and the other particles is sufficiently high. A standard test for such a model is to study the duration it takes to sink a SMBH into the centre of a system (see e.g. \citealt{Tremmel_subgrid_2015, Damiano2025}). In this section, we study how accurately \ketju{} and the two dynamical friction subgrid models discussed in Section \ref{sect:dyn_fric_subgrid} follow the analytical estimate for a SMBH sinking in an isolated system. Note that the analytical expression breaks down at small separations from the centre (see Section \ref{sect:plummer_ics_and_analytical}).

\subsection{Initial conditions}
\label{sect:plummer_ics_and_analytical}

Following \cite{Gadget4-ketju} we start by studying the orbital decay of a SMBH due to dynamical friction in a system consisting of a stellar component following a Plummer \citep{1911MNRAS..71..460P} density profile. For a Plummer sphere, the density profile with total mass $M$ and scale radius $a_{\rm s}$ is
\begin{equation}
    \rho(r) = \frac{3M}{4\pi a_{\rm s}^3}\left(1+\left( \frac{r}{a_{\rm s}} \right)^2 \right)^{-5/2}.
\end{equation}
The stellar mass of the system is $M=10^{11}\ \mathrm{M}_\odot$ and the scale radius is set at $a_{\rm s}=1.5\ \mathrm{kpc}$. The SMBH of mass $m_\bullet=3\times 10^7\,\mathrm{M}_\odot$ (this mass is close to the $m_\mathrm{Ketju}$ limit, which we will use in the cosmological zoom-in simulation) is originally set to be on a circular orbit at a distance equal to the virial radius of the Plummer sphere, $r_\mathrm{vir}=16\left(\frac{a_{\rm s}}{3\pi}\right)\approx 2.55\,\mathrm{kpc}$ from the centre of the system.

For this density profile, the analytic expectation for the orbital decay is derived in \citet{Rodriguez2018}. The rate of change for the distance from the centre is given by
\begin{equation}
    \frac{\mathrm{d} r}{\mathrm{d} t} = - \frac{8\pi G^2\ln \Lambda \rho(r)\chi m_\bullet r}{v_c^3\left( 
1+3(1+r^2a_{\rm s}^{-2})^{-1} \right)},
\end{equation}
where $\ln\Lambda$ is the Coulomb logarithm,
\begin{equation}
    v_c=\sqrt{\frac{GM}{r}} \left(1+\frac{a_{\rm s}^2}{r} \right)^{-3/4}
\end{equation}
is the circular velocity and 
\begin{equation}
    \chi = \mathrm{erf}(X)-\frac{2}{\sqrt{\pi}}X\exp\left(-X^2\right),
\end{equation}
where
\begin{equation}
    X=\frac{v_c}{\sqrt{2}\sigma(r)}.
\end{equation}
Here, $\sigma$ is the velocity dispersion
\begin{equation}
    \sigma(r) = \sqrt{\frac{GM}{6\sqrt{a_{\rm s}^2+r^2}}},
\end{equation}
and with our parameters the Coulomb logarithm can be estimated to be (e.g. \citealt{BinneyTremaine})
\begin{equation}
    \ln\Lambda=\ln\left( \frac{b_\mathrm{max}}{b_{90}} \right)=\ln\left(\frac{2r_\mathrm{vir}}{Gm_\bullet\sigma(r_\mathrm{vir})^{-2}} \right)\approx 5.7.
\end{equation}

It is known from previous studies that for small separations from the centre, the analytic formula breaks down, reaching zero separation which is not seen in simulations (e.g. \citealt{2006MNRAS.368.1073G, 2021ApJ...916....9M, Gadget4-ketju}). This phenomenon is known as core stalling and has been studied in various works using either $N$-body simulations \citep{2010ApJ...725.1707G, 2011MNRAS.416.1181I} or an analytic approach \citep{2018ApJ...868..134K}. Recently, \citet{2025arXiv251111804D} showed that core stalling is caused by the  specific phase-space structure of the galaxy, with core stalling occurring when the distribution function (which describes the phase-space structure) reaches a plateau, resulting in a zero net torque. 

\subsection{Black hole sinking}
\label{sect:sinking}

We simulate the system using four different models: \textsc{gadget-3}, \ketju{} and the two dynamical friction subgrid models of \cite{Tremmel_subgrid_2015} and \cite{Ma_discrete_subgrid_2023}. With each model, the SMBH orbital decay is simulated starting from five different random realisations of the same underlying stellar phase space
sampling. The softening length for all particles is $\epsilon=50\,\mathrm{pc}$, resulting in a \ketju{} radius of $r_\mathrm{Ketju}=150\,\mathrm{pc}$. BH repositioning is not used in any of the simulations. 

Figures \ref{fig:plummer_100} and \ref{fig:plummer_10} show results from the simulations with BH to stellar particle mass ratios of $m_\bullet/m_\star=100$ and $m_\bullet/m_\star=10$, respectively. For all four dynamics models, the larger panel shows the evolution of the SMBH separation $r$ from the centre of the sphere (coloured lines) and the analytical estimate (black line), while the smaller panel shows the relative difference in the evolution of the SMBH distance from the centre of the system compared to the analytic estimate. 
The simulations using only \textsc{gadget-3} fail to sink the SMBH to the centre for both particle mass ratios, with the SMBH reaching a $1\,\mathrm{kpc}$ separation from the centre more than $1\,\mathrm{Gyr}$ later than the analytic estimate in the $m_\bullet/m_\star=100$ runs. In simulations with the lower resolution $(m_\bullet/m_\star=10)$, one simulation (Run D) fails to reach a separation smaller than $1.5\,\mathrm{kpc}$ during the $3.5\,\mathrm{Gyr}$ runtime, signifying the need for an improved model for SMBH dynamics. 

The analytical estimate is best matched with simulations using either \ketju{} or the \cite{Tremmel_subgrid_2015} subgrid model, especially in the $m_\bullet/m_\star=100$ mass ratio runs. In these runs, at the end of the simulations, the SMBH distance from the centre only slightly deviates from the analytical prediction. While the analytical estimate reaches zero separation, the SMBH stalls when it reaches a distance of $\sim 150\,\mathrm{pc}$ from the centre. As discussed in Section \ref{sect:plummer_ics_and_analytical}, the discrepancy is caused by core stalling, which is not captured in the analytic expression. Focusing on the lower mass resolution, the \ketju{} and \cite{Tremmel_subgrid_2015} models still follow the analytical estimate when the mass ratio is decreased to $m_\bullet/m_\star=10$, but the deviations from the analytical estimate reach larger values (the largest residual is $1.0\,\mathrm{kpc}$ for \ketju{} and $0.5\,\mathrm{kpc}$ for the subgrid model). 

At the end of the mass ratio $m_\bullet/m_\star=10$ runs, the \ketju{} simulations have the SMBH oscillating at a distance of around $500\,\mathrm{pc}$ from the centre and in the subgrid model of \cite{Tremmel_subgrid_2015} the SMBHs oscillate at distance of around $300\,\mathrm{pc}$. The larger offsets for \ketju{} at this low mass ratio are due to individual stellar particles being massive enough to noticeably affect the trajectory of the SMBH in hard scattering events. Nonetheless, \ketju{} is still in a relatively good agreement with the analytical estimate with the small mass ratio of $m_\bullet/m_\star=10$, which is a significantly lower mass resolution than what previous studies have found necessary \citep{Gadget4-ketju}. However, we note that previous studies with \ketju{} have focused on the binary phase instead of dynamical friction.

With the discrete dynamical friction model, the simulations with mass ratio $m_\bullet/m_\star=10$ follow the analytical sinking estimate better compared to the simulations with a mass ratio of 100. In addition, core stalling is seen only with the smaller mass ratio. With increased resolution, the SMBH sinks too rapidly and reaches a separation of $500\ \mathrm{pc}$ from the centre of the system around $500 \ \mathrm{Myr}$ before the analytic estimate. After reaching the centre, the subgrid model very efficiently keeps the SMBH at the centre. The too rapid sinking is due to the so-called `double counting' of gravity, inherent to this subgrid model. Since a cosmological environment will contain SMBHs with different masses, i.e. different mass ratios $m_\bullet/m_\star$ and since we want to avoid `double counting' gravitational force for all SMBH masses, we adopt the \cite{Tremmel_subgrid_2015} model as the fiducial subgrid dynamical friction model in \ketju{}.

\section{Brownian motion in a Hernquist sphere}
\label{sect:brownian}

In addition to SMBH sinking, a model for SMBH dynamics must be able to keep SMBHs at the centres of their host systems. Here, we therefore study the Brownian motion (e.g. \citealt{Merritt2001,2016MNRAS.461.1023B}) for SMBHs of various masses put at the centre of an isolated system. In a cosmological environment, this is crucial for the mass growth of SMBHs, as the inability to effectively keep SMBHs in the central regions of galaxies can result in negligible mass growth \citep{2022MNRAS.516..167B}. The Brownian motion of the SMBHs is also important for GW sources, as the Brownian motion of the SMBHs can lead to changes in the hardening rate \citep{2016MNRAS.461.1023B}, potentially affecting the predictions of the GW merger rate. 

Before the study of Brownian motion, we present a modification for choosing the size of the \ketju{} integrated region, which is not only used in this section but also in the cosmological zoom-in simulations (Sections \ref{sect:gal_evo} and \ref{sect:cosmo_bh}). This modification introduces smaller softening lengths for SMBHs than stellar particles (which before was not possible with \ketju{}) in order to keep the number of particles within the \ketju{} radius $(r_\mathrm{Ketju})$ of a SMBH from reaching an excessively high values.

\label{sect:Brownian}

\subsection{Size of \textsc{ketju} integration region}

With \textsc{ketju}, the dynamical interactions between SMBHs and all stellar particles are non-softened. Previously, the softening lengths of these two particles types have been tied to the \textsc{ketju} radius via the relation $r_\mathrm{Ketju}\geq 2.8 \epsilon_{\bullet\leftrightarrow\star}$, where $\epsilon_{\bullet\leftrightarrow\star}$ is the softening length used for the interactions between SMBHs and stellar particles. In standard \textsc{gadget-3} the chosen value is the maximum of the softening lengths of the two particle types, i.e. $\epsilon_{\bullet\leftrightarrow\star}=\mathrm{max}(\epsilon_\bullet,\ \epsilon_\star)$. 

Due to the relatively steep scaling of \ketju{} with the number of particles in a \ketju{} region (roughly $\mathcal{O}(N^2)$, see \citealt{MSTAR}), the computational cost can become unsustainable, especially in dense environments unless the softening lengths of stars are set to small values of the order of a few pc. To avoid putting the softening of stellar particles to such small values, we instead decrease the BH softening length and limit the \ketju{} region radius only with the BH softening length, i.e. $r_\mathrm{Ketju}\geq 2.8\epsilon_\bullet$. In order to keep every BH--stellar particle interaction non-softened, the chosen softening length for interactions between BHs and stellar particles in \textsc{gadget-3} is now $\epsilon_{\bullet\leftrightarrow\star}=\mathrm{min}(\epsilon_\bullet,\ \epsilon_\star)=\epsilon_\bullet$. For all other particle pairs, the softening length used is still the maximum softening length of the two particle types. With this modification, only the softening length of BHs needs to be changed to make the simulation of dense systems computationally feasible. The change is used in all presented simulations where stellar particles and BHs have unequal softening lengths.

The presented change to the softening length of SMBHs does not affect the dynamics of single SMBHs. We show that in both isolated and cosmological simulations the \ketju{} integrated SMBHs remain at the centres of their host galaxies (see \autoref{subsect:brownian} and \autoref{subsect:cosmo_bh_dyn}). 
As PN terms for SMBH binaries are included in the force calculations only when two \ketju{} regions overlap, it is also important for these terms to be negligible when the separation between two \ketju{} integrated SMBHs becomes $<2\times r_\mathrm{Ketju}$. Since the \ketju{} radius is set to $r_\mathrm{Ketju}\sim 15\ \mathrm{pc}$ in this study, the binary evolution is always dominated by Newtonian acceleration when the regions begin to overlap.

\begin{figure}
    \centering
    \includegraphics[width=\linewidth]{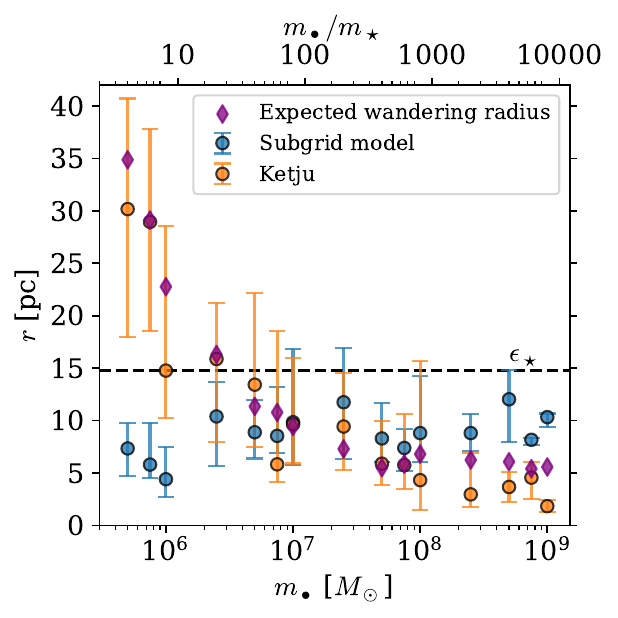}
    \caption{The median distance of the SMBH from the centre for various SMBH masses, with errorbars covering the range from 25th percentile to 75th percentile. The results are shown for a set of simulations run with \ketju{} and for a set where the dynamical friction subgrid model is used. For \ketju{} integrated SMBHs, the Brownian motion starts to increase as the  mass ratio falls below $m_\bullet/m_\star\lesssim 100$. With the subgrid model, the SMBH remains within the stellar softening $\epsilon_\star$ (shown as a dashed line) at all BH masses.}
    \label{fig:brownian_motion}
\end{figure}

\subsection{Initial conditions}

The motion of SMBHs integrated using \ketju{} and its dependence on the $m_\bullet/m_\star$ mass ratio is studied by investigating the wandering radius of a SMBH in a cuspy density profile. Here, an SMBH with zero initial velocity is put at the centre of the system consisting of a stellar and dark matter component. Both components follow the Hernquist density profile \citep{Hernquist1990}. With this, the density $\rho$ of component $i$ at radius $r$ is 
\begin{equation}
    \rho(r) = \frac{M_i}{2\pi}\frac{a_i}{r(a_i+r)^3},
\end{equation}
where $M_i$ and $a_i$ are the total mass and the scale radius of the component, respectively. The mass of the stellar component is set by using the $m_\bullet-M_\star$ relation from \cite{Kormendy2013} with $m_\bullet=10^8\,\mathrm{M}_\odot$. After this, the dark matter halo mass is calculated using the stellar-to-halo-mass relation from \cite{Moster2013}. The effective half-mass radius of the stellar component is set to $R_\mathrm{e}=1\,\mathrm{kpc}$. Similar to \cite{rantala2018}, we set the dark matter mass fraction within the stellar half-mass radius to 25\%. Putting all of these together, the stellar component of the isolated galaxy has a mass of $M_\star \approx 2.5\times 10^{10}\,\mathrm{M}_\odot$ and a scale radius of $a_\star\approx 2.2\,\mathrm{kpc}$, while the dark matter halo has a mass of $M_\mathrm{dm} \approx 6.8\times 10^{11}\,\mathrm{M}_\odot$ and a scale radius of $a_\mathrm{dm}\approx 62\,\mathrm{kpc}$. 

The system is simulated with various different SMBH masses. For dark matter and baryonic particles (here just stellar particles), the mass ratio is set to $m_\mathrm{dm}/m_\star=\Omega_\mathrm{dm}/\Omega_\mathrm{b}=0.266/0.049\approx 5.4$, using values from \cite{Planck}. The particle masses are $m_\star\approx 1.2\times10^5\,\mathrm{M}_\odot$ and $m_\mathrm{dm}\approx 6.72\times10^5\,\mathrm{M}_\odot$, respectively. The stellar and dark matter particle masses are equal to the resolution of our zoom-in cosmological simulation (see Section \ref{sect:gal_evo}). In total 15 different masses are used for the SMBH at the centre of the system, varying in the range $m_\bullet=5\times 10^5\,-10^9\,\mathrm{M}_\odot$. 

Each simulation in this set is run for $2\,\mathrm{Gyr}$. Two versions are run, one using \ketju{} and the other with the dynamical friction subgrid model of \cite{Tremmel_subgrid_2015}. As is the case with the particle masses, the softening lengths are chosen to be equal to the ones used in the zoom-in simulations after $z=9$. The softening length for dark matter is set to $\epsilon_\mathrm{dm}=42\,\mathrm{pc}/h=62.3\,\mathrm{pc}$ ($h=0.674$), while the stellar particle and SMBH softening lengths are set to $\epsilon_{\star}=10\,\mathrm{pc}/h=14.8\,\mathrm{pc}$ and $\epsilon_{\bullet}=3.3\,\mathrm{pc}/h=4.9\,\mathrm{pc}$ ($r_\mathrm{Ketju}=3\times \epsilon_{\bullet}=10\,\mathrm{pc}/h$), respectively.

\subsection{Brownian motion}
\label{subsect:brownian}

For a SMBH experiencing Brownian motion, the mean square velocity $\langle v^2_\bullet \rangle$ and mean square displacement $\langle r_\bullet^2 \rangle$ are related via (assuming constant density, \citealt{Merritt2001})
\begin{equation}
    \langle v_\bullet^2 \rangle = \frac{4}{3}\pi G \rho_f \langle r_\bullet^2 \rangle,
\end{equation}
where $\rho_f$ is the density of the field particles. We will here assume that the Brownian motion is only due to the stellar component of the system (inside the central $100\ \mathrm{pc}$, the mass fraction $m_\star/(m_\star+m_\mathrm{dm})\gtrsim 0.95$). Since $\langle v^2_\bullet \rangle$ can be written using the 1D velocity dispersion of the stellar component as (e.g. \citealt{Merritt2007})
\begin{equation}
    \langle v_\bullet^2\rangle = 3 \frac{m_\star}{m_\bullet}\sigma_\star^2,
\end{equation}
the wandering radius (root mean square of displacement) can therefore be written as
\begin{equation}
    r_\mathrm{wander} = \sqrt{\langle r_\bullet^2 \rangle} = \frac{3}{2} \sqrt{ \frac{1}{\pi G \rho_\star} \frac{m_\star}{m_\bullet} \sigma^2_\star}.
\end{equation}
For each simulation, $\sigma_\star$ and $\rho_\star$ are calculated for each snapshot (saved in intervals of $100\ \mathrm{Myr}$) using star particles that are within a distance of $\epsilon_\star$ from the centre.

While the accuracy of \ketju{} in terms of sinking an SMBH with a mass ratio as low as $m_\bullet/m_\star=10$ is surprisingly close to the dynamical frction subgrid model (Figure \ref{fig:plummer_10}), the two models show clear differences at small mass ratios when studying the Brownian motion of SMBHs. In Figure \ref{fig:brownian_motion} the median separations of SMBHs with various masses are shown. The lower limit for the errorbars is the 25th percentile and the upper limit is the 75th percentile from simulations with runtimes of $2\,\mathrm{Gyr}$. For SMBHs with masses $>10^7\,\mathrm{M}_\odot$ ($m_\bullet/m_\star\gtrsim 100$), the median distance from the centre remains below $10\,\mathrm{pc}$ for \ketju{} integrated SMBHs. As the SMBH mass is decreased to $m_\bullet<10^7\,\mathrm{M}_\odot$, the SMBHs in the \ketju{} simulations wander farther from the centre, and at SMBH masses of 
$\sim 10^6\,\mathrm{M}_\odot$ (roughly the seed mass for the zoom-in simulations in Section \ref{sect:gal_evo}) the median distance sharply rises to tens of pc.

The expected wandering radii (the mean of the calculated radius from all snapshots) for each simulation is shown as purple diamonds in Figure \ref{fig:brownian_motion}. Results from the \ketju{} integrated simulations (orange circles) and the calculated expectation for the wandering radius (purple diamonds) follow the same trend, with the expected wandering radius remaining within the shown errorbar at all mass ratios except when the ratio is above a few thousand. For the largest mass ratios, the expected wandering radius is a few pc above the errorbar.

For the tested SMBH masses, the Brownian motion of the dynamical subgrid model of \cite{Tremmel_subgrid_2015} does not show a strong dependence on the mass ratio. The median distance from the centre is below the stellar softening length $\epsilon_\star$ (black dashed horizontal line in the figure) for all masses and the 75th percentile remains at less than $20\,\mathrm{pc}$. The weak dependence of the mass ratio on the Brownian motion is not surprising, since the tested mass ratios are roughly equal to or slightly larger than the resolutions tested in \cite{Tremmel_subgrid_2015}. Comparing the median displacements of SMBHs between the two simulations, \ketju{} reaches slightly smaller values compared to the subgrid model as the mass ratios increase above $m_\bullet/m_\star>100$.

\begin{table*}
\begin{tabular}{|c | c|c|c|c|c|c|c|c|c|c|c}
\hline
ID &  $z_{\bullet,\mathrm{seed}}$ & $z_{\bullet,\mathrm{Ketju}}$ &$z_{\bullet,f}$ & $M_{\star,f}$& $m_{\bullet,f}$ & $m_{\bullet,f}/M_{\star,f}$ & $r_{\mathrm{eff},f}$ & min$(r_\mathrm{eff})$ & $z_\mathrm{compact}$ & $f_\mathrm{in-situ}(>z_\mathrm{compact})$ \\
 &  &  & &  [$10^{10}\ \mathrm{M}_\odot$] & [$10^{7}\ \mathrm{M}_\odot$]& &  [pc]& [pc]& &  \\
\hline
A & 14.04 & 7.00 & 5.00 & 8.509 & 31.954 & 0.0038 & 305.4 & 124.6 & 9.14 & 0.87 \\
B($\rightarrow$A) & 12.88 & 7.54 & 5.22 & 8.313 & 14.677 & 0.0018 & 274.9 & 49.0 & 6.80 & 0.94 \\
C& 12.01 & 6.31 & 5.00 & 1.224 & 4.021 & 0.0033 & 456.1 & 84.9 & 7.58 & 0.94 \\
D($\rightarrow$A) & 11.94 & 6.64 & 6.25 & 3.324 & 2.798 & 0.0008 & 207.5 & 39.6 & 8.13 & 0.93 \\
E & 11.56 & 6.75 & 5.00 & 2.056 & 7.454 & 0.0036 & 274.6 & 143.2 & 8.77 & 0.87 \\
F & 10.95 & 6.11 & 5.00 & 0.927 & 6.040 & 0.0065 & 64.1 & 28.1 & 7.85 & 0.95 \\
G & 9.60 & 5.60 & 5.00 & 2.276 & 6.171 & 0.0027 & 227.2 & 88.0 & 7.33 & 0.65 \\
H & 9.60 & 5.13 & 5.00 & 0.710 & 1.955 & 0.0028 & 260.9 & 97.8 & 6.25 & 0.91 \\
I & 8.59 & 5.22 & 5.00 & 0.460 & 2.782 & 0.0061 & 39.4 & 38.1 & 5.79 & 0.93 \\
J($\rightarrow$G) & 9.87 & - & 5.16 & 2.033 & 1.282 & 0.0006 & 216.1 & 131.7 & 6.33 & 0.87 \\
\hline
\end{tabular}
\caption{Properties of galaxies and \ketju{} integrated SMBHs derived from the simulation zoom-K+DF. Right arrow pointing to another ID is included in the first column if a SMBH merges with a larger SMBH. $z_{\bullet,\mathrm{seed}}$ is the redshift at the moment when the SMBH was seeded and $z_{\bullet,\mathrm{Ketju}}$ is the redshift at which it reached the $m_\mathrm{Ketju}$ mass. $z_{\bullet,\mathrm{Ketju}}$ is not shown for SMBH J, since it merges with G before reaching this mass limit. $z_{\bullet,f}$ is the redshift of the final snapshot where the SMBH existed. This is equal to $z=5$, unless the SMBH merged with another SMBH. $M_{\star,f}$, $m_{\bullet,f}$, $m_{\bullet,f}/M_{\star,f}$ and $r_{\mathrm{eff},f}$ give the stellar mass, SMBH mass, the mass ratio of the SMBH and the stellar component and the effective stellar radius at redshift $z_f$, respectively. Finally, min$(r_\mathrm{eff})$ is the minimum effective radius the galaxy reaches during the simulation and $z_\mathrm{compact}$ is the redshift at which this occurs, while $f_\mathrm{in-situ}(>z_\mathrm{compact})$ is the fraction of stars formed in-situ before $z_\mathrm{compact}$.}
\label{table:properties}
\centering
\end{table*}

\section{Cosmological zoom-in simulations of high-redshift galaxies}
\label{sect:gal_evo}

Next, we study the evolution of galaxies and the dynamics of SMBHs in a cosmological setting. By performing a cosmological zoom-in simulation, we target a dense region centred on a dark matter halo with a virial mass of $M_\mathrm{vir}\approx 1.7\times 10^{12}\,\mathrm{M_\odot}$ at redshift $z=5$. The virial mass is calculated as the sum of dark matter mass within a virial radius $r_\mathrm{vir}$, which is the radius at which the density is 200 times the critical density of the Universe. 
In addition, utilising the model for dynamics presented in Section \ref{sect:K+T}, we can study the dynamics of SMBHs starting from their seeding all the way to the GW-driven evolution of SMBH binaries. This is the first simulation which resolves the entire SMBH dynamics evolution in a single simulation.

We perform three cosmological zoom-in simulations, each focusing on the same volume but with a different implementation for SMBH dynamics. One of the simulations is run with the new hybrid model that combines \ketju{} and the dynamical friction subgrid model (henceforth named zoom-K+DF). As discussed in Section \ref{sect:K+T}, the \ketju{} integrated SMBH binaries coalesce when a separation smaller than twelve times their summed Schwarzschild radius is reached. Lower mass non-\ketju{} binaries are merged when two SMBHs are within the SPH smoothing length $h_{\rm SPH}$ with the relative velocity being below half of the local sound speed. For the other two simulations, all SMBH mergers are treated this way. The second run (named zoom-DF) is otherwise identical to the zoom-K+DF simulation except for the fact that the \ketju{} integration is disabled, i.e. the simulation is performed using just the dynamical friction subgrid model. The third simulation (named zoom-G) is run with standard \textsc{gadget-3} and uses BH repositioning (e.g. \citealt{2005MNRAS.364.1105S, Johansson2009}), meaning that at each time the position of the BH particle is updated, it is forced to be at the local minimum of the gravitational potential.

This section studies the evolution of the galaxies with \ketju{} integrated SMBHs. We focus on the stellar component of the galaxies, and how it co-evolves with the SMBHs. 
The stellar properties are calculated from stellar particles within the galaxy radius defined as $r_\mathrm{gal}=0.1r_\mathrm{vir}$. Apart from Figure \ref{fig:bh-stellar-relation}, results are only shown from the zoom-K+DF simulation.

\subsection{Initial conditions}

The zoom-in volume has a size of $(4.5\,\mathrm{cMpc}/h)^3$ containing $\sim 2\times 220^3$ particles, while the total volume of the simulation box is $(75\,\mathrm{cMpc}/h)^3$. The initial conditions were generated using the \textsc{music} software \citep{Hahn2011}. We use the cosmological parameters from \citet{Planck}, with 
$\Omega_\mathrm{m}=0.315$, $\Omega_\mathrm{b}=0.0491$, $\Omega_\Lambda=0.685$ and $h=0.674$. The particle masses in the zoom-in region are $m_\mathrm{bar}=8.4 \times 10^4\,\mathrm{M}_\odot/h$ for baryonic particles (i.e. gas and stars, note that the supernova feedback model can slightly alter their masses) and $m_\mathrm{dm}=4.5 \times 10^5\,\mathrm{M}_\odot/h$ for dark matter particles. For the simulation where \ketju{} is enabled, the mass limit for \ketju{} integration is $m_\mathrm{Ketju}=10^7\,\mathrm{M}_\odot/h$, equal to the mass ratio $m_\mathrm{Ketju}/m_\star\approx120$, which was found to be sufficient to keep a SMBH at the centre of its host galaxy (see Section \ref{sect:brownian}). 

SMBHs with masses of $m_\mathrm{seed}=10^{6}\ \mathrm{M}_\odot/h$ are seeded into halos found using the Friends-of-Friends algorithm \citep{1985ApJ...292..371D} as halos reach a mass of $5\times10^{9}\ \mathrm{M}_\odot/h$, the same halo mass threshold that was used in the \textsc{Astrid} simulation \citep{astrid, astrid_smbhs}. The SMBH seed mass is relatively large, but still roughly equal to the seed mass used in the ROMULUS \citep{Romulus} and the IllustrisTNG simulations \citep{2018MNRAS.479.4056W}. The gravitational softening lengths after reaching redshift $z=9$ are $\epsilon_\mathrm{gas}=20\,\mathrm{pc}/h$, $\epsilon_\star=10\,\mathrm{pc}/h$, $\epsilon_\bullet=3.3\,\mathrm{pc}/h$ (resulting in $r_\mathrm{Ketju}=10\,\mathrm{pc}/h$) and $\epsilon_\mathrm{dm}=42\,\mathrm{pc}/h$ for gas, stellar, SMBH and high resolution dark matter particles, respectively. The softening length for the low resolution dark matter is $\epsilon_\mathrm{low}=5.96\,\mathrm{kpc}/h$. Before $z=9$, the softening lengths are fixed in comoving coordinates and after $z=9$ they remain fixed in physical coordinates.

\subsection{Growth of galaxy masses and sizes}

\begin{figure*}
    \centering
    \includegraphics[width=0.8\linewidth]{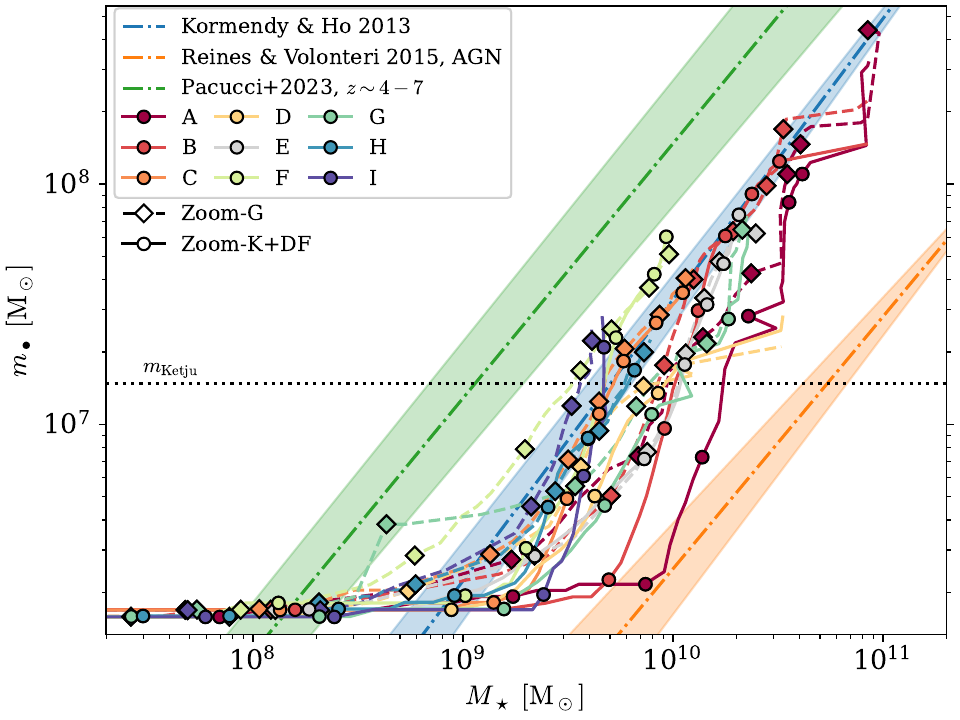}
    \caption{The evolution of the SMBH mass as a function of the stellar mass for each system that reaches $m_\bullet\geq m_\mathrm{Ketju}$. The diamonds and dashed lines show the evolution from the simulation zoom-G, while the circles and solid lines represent the simulation zoom-K+DF. The mass limit of switching \ketju{} integration on for a SMBH is shown as a horizontal dotted line. Markers are shown for each system at intervals of $\sim 105\ \mathrm{Myr}$. The blue and orange dash-dotted lines show the local relations from \citet{Kormendy2013} and \citet{Reines2015}, respectively. The green dash-dotted line is the relation calculated by \citet{Pacucci23} using observations with JWST of SMBHs at $z\sim 4-7$. The $1\sigma$ error for each relation is included. Simulated systems start to follow the local relation as soon as the SMBHs begin to grow efficiently. In simulations without BH repositioning the SMBH seeds start growing at somewhat higher galaxy stellar masses.
    }
    \label{fig:bh-stellar-relation}
\end{figure*}

In Figure \ref{fig:bh-stellar-relation} we show the evolution of the galaxy stellar mass, compared against the mass of the SMBH for the simulations where \ketju{} was enabled (zoom-K+DF, circles) and from the simulation where repositioning was instead used (zoom-G, diamonds).
For each line, the time interval between two markers is around $105\ \mathrm{Myr}$. Throughout sections \ref{sect:gal_evo} and \ref{sect:cosmo_bh}, we focus on systems which have a SMBH reaching mass $m_\mathrm{Ketju}$ during the simulation. The systems are named as letters from A to I. Dash-dotted lines in the background represent observed $m_\bullet-M_\star$ relations, with the \cite{Kormendy2013, Reines2015} relation being based on observations from the low redshift Universe, while the \cite{Pacucci23} relation is derived from JWST observations of overmassive SMBHs at high redshifts of $z\sim 4-7$.

In the simulation without repositioning, the start of the SMBH growth is delayed. When repositioning is switched on, the SMBH mass accretion starts when the stellar mass of the galaxy is a few times $10^8\,\mathrm{M}_\odot$, while the growth starts only after the stellar component has a mass above $10^9\,\mathrm{M}_\odot$ when the dynamics of SMBHs is modelled with the dynamical friction subgrid model. The delay, which is physical, is caused by the SMBHs not being immediately at the centres of their host galaxies when the density becomes sufficiently high for effective SMBH accretion, since dynamical friction needs to sink the SMBHs to the centres (see also Figure \ref{fig:displacements}). Although the start of the SMBH growth occurs at different stellar masses, the SMBHs from both simulations closely follow the relation of \cite{Kormendy2013} especially after the SMBHs reach the \ketju{} mass limit $(m_{\bullet}\sim m_\mathrm{Ketju})$.

\begin{figure*}
    \centering
    \includegraphics[width=\linewidth]{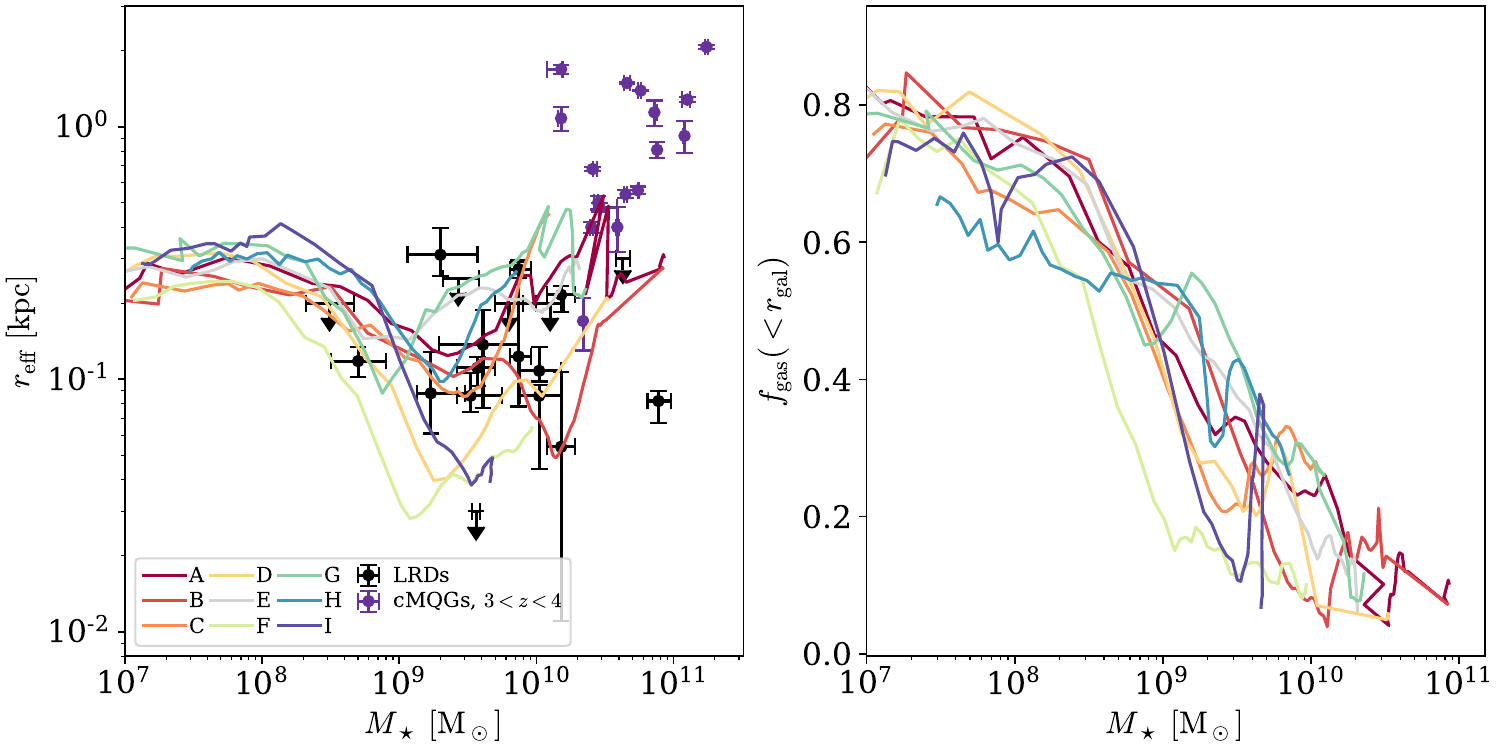}
    \caption{Left: The evolution of the stellar mass and the effective radius of the stellar component of each system which reaches $m_\bullet\geq m_\mathrm{Ketju}$ in the simulation zoom-K+DF. The black dots show the estimated sizes and masses of LRDs observed with JWST \citep{Baggen23, Kokorev24, Wang25, Akins25, Ma2025_lrd, Baggen2024, Akins2023}. For observations where only a upper limit was available for the size, an arrow indicates the upper limit of the size. The purple points show sizes and masses of compact massive quiescent galaxies in the redshift range $3<z<4$, also observed using JWST \citep{Kawinwanichakij25}. Galaxies grow through a phase of compaction, starting roughly when $M_\star = 10^8\,\mathrm{M}_\odot$. Right: the evolution of the gas fraction within $r_\mathrm{gal}=r_\mathrm{vir}/10$ as a function of the stellar mass. The gas fraction of each galaxy begins to decrease around the same time when the phase of compaction begins
    at stellar masses of $M_{\star}\sim 10^{8.5}-10^{9} \ M_{\odot}$. 
    }
    \label{fig:size-evolution}
\end{figure*}

The size evolution of the stellar component, as a function of the stellar mass, of the galaxies whose SMBHs reach the \ketju{} integration limit in the simulation zoom-K+DF is shown in the left panel of Figure \ref{fig:size-evolution}. The size here is measured by the effective radius $r_\mathrm{eff}$, which is defined as the mean projected stellar half-mass radius, as viewed from  60 random viewing angles. Sizes and masses of LRDs observed with JWST are taken from various sources \citep{Baggen23, Kokorev24, Wang25, Akins25, Ma2025_lrd, Baggen2024, Akins2023} and are shown as black markers, while the purple markers represent compact massive quiescent galaxies (cMQGs), also observed by JWST but at slightly lower redshifts (\citealt{Kawinwanichakij25}, $3<z<4$). The minimum size $(\rm min(r_{\rm eff}))$ and the final size $(r_{\rm eff,f}$, i.e. the size at the final redshift $z_{\bullet,f}$ at which the SMBH exists) are shown for each system in \autoref{table:properties}.

All galaxies that manage to efficiently grow their SMBH above the \ketju{} integration limit before $z=5$ go through a phase of compaction. The sizes first slightly increase with the growing stellar mass, reaching effective radii of a few hundred pc. Around stellar masses of $M_{\star}\gtrsim 10^8\,\mathrm{M}_\odot$ the sizes begin to shrink while the galaxies grow in mass, with the most compact sizes being $30-140\,\mathrm{pc}$, depending on the galaxy. In \autoref{table:properties} we show the fraction $f_\mathrm{in-situ}(>z_\mathrm{compact})$ of stars formed in-situ before redshift $z_\mathrm{compact}$. The fraction is calculated by looping backwards from the snapshot at $z_\mathrm{compact}$. For each stellar particle, we take the snapshot that is nearest to the time at which a gas particle has changed to the stellar particle. If the particle at this snapshot is within $r_\mathrm{gal}$, it is considered to be formed in-situ. Star formation is heavily dominated by in-situ formation as the fraction is generally above $90\%$ for all galaxies, meaning that the compaction is caused by centrally concentrated star formation. The compact sizes also agree well with the observed sizes of LRDs. The phase of compaction lasts from roughly $100\ \mathrm{Myr}$ to $200\ \mathrm{Myr}$, after which the sizes begin to increase again due to off-centre star formation and the growth continues until the simulations stop at a redshift of $z=5$. At this point, the sizes and masses are starting to overlap with the population of cMQGs. We further discuss the evolution of our simulated galaxies from masses and sizes matching observations of LRDs to larger systems in Section \ref{sect:discussion_highz_gals}.

The right panel of Figure \ref{fig:size-evolution} shows the evolution of the baryonic gas fraction $f_\mathrm{gas}=M_\mathrm{gas}/(M_\star+M_\mathrm{gas})$ as a function of stellar mass, calculated within the radius $r_\mathrm{gal}$. The gas fraction of all shown galaxies follow the same trend, where the gas fraction is initially high, around 80\%. Around the same stellar mass $(M_{\star}\sim 10^{8.5}-10^{9} \ M_{\odot})$ as the stellar size decreases, the gas fraction rapidly begins to decrease. The resulting gas fraction varies between different galaxies with the values being in general between 10\% and 40\%. 

\begin{figure*}
    \centering
    \includegraphics[width=0.9\linewidth]{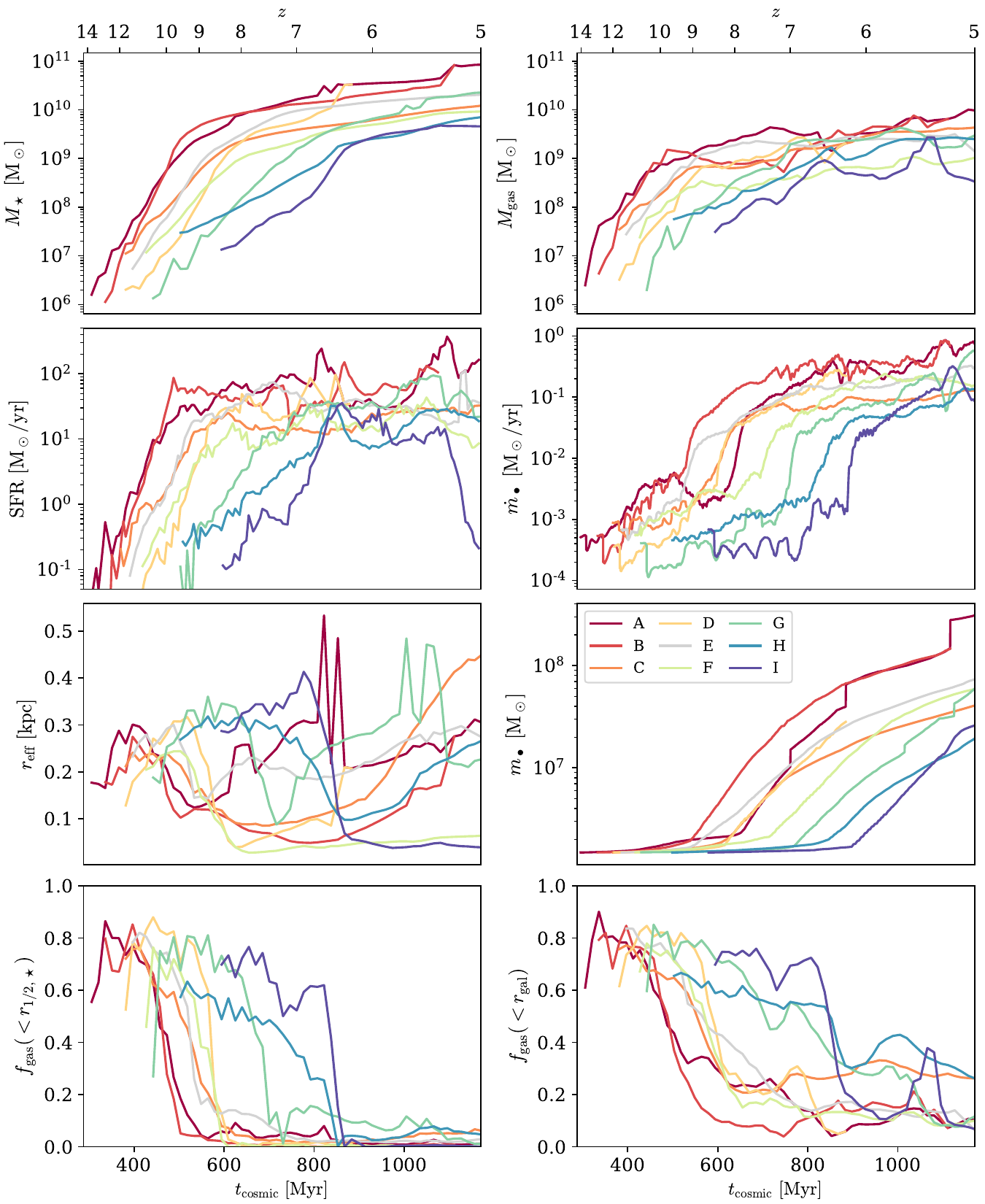}
    \caption{Time evolution of eight properties for each system which reaches $m_\bullet\geq m_\mathrm{Ketju}$ in the zoom-K+DF simulation. The left column, from top to bottom, shows the stellar mass of the galaxy, the star formation rate, the effective radius of the stellar component and the gas fraction $f_{\rm gas}=M_\mathrm{gas}/(M_\mathrm{gas}+M_\star)$ calculated within the three-dimensional stellar half-mass radius, respectively. The right column, from top to bottom, shows the gas mass of the system, the SMBH accretion rate, the SMBH mass and the gas fraction within the radius $r_\mathrm{gal}=r_\mathrm{vir}/10$, respectively.
    Within the half-mass radius, the gas fraction experiences a sharp drop occurring roughly at the same time the effective radius decreases. At larger radii a drop in the gas fraction is also seen, but the system is not fully depleted from gas. For each galaxy, the total gas mass is not significantly reduced, although the stellar mass is also increasing due to in-situ star formation, meaning that there is a net gas inflow into every galaxy.
    }
    \label{fig:multipanel}
\end{figure*}

\begin{figure*}
    \centering
    \includegraphics[width=0.8\linewidth]{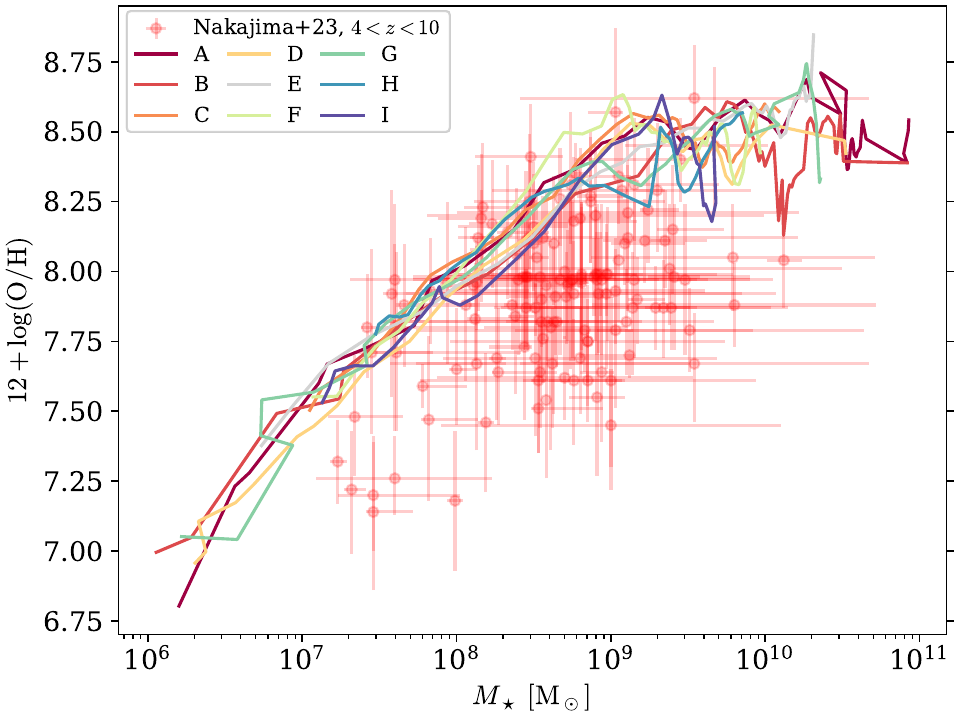}
    \caption{The evolution of the stellar mass and the stellar metallicity, as measured by $12+$O/H of each system which reaches $m_\bullet\geq m_\mathrm{Ketju}$ in the zoom-K+DF simulation. The red markers show JWST observations in the redshift range $4<z<10$ \citep{Nakajima2023}. The evolution of metallicity matches observations throughout the mass range $ 10^7 - 10^{10}\ \mathrm{M}_\odot$.
    }
    \label{fig:metallicity}
\end{figure*}

In order to take a closer look on the evolution of the galaxies during and after the phase of compaction, Figure \ref{fig:multipanel} presents the evolution of eight different parameters as a function of time from the zoom-K+DF simulation. Each line starts from the moment when the SMBH was seeded. The top row includes the evolution of the galaxy stellar mass (left panel) and the galaxy gas mass (right panel). Both are calculated within the radius $r_\mathrm{gal}$. Although the gas mass includes momentary drops, the gas mass in general steadily increases as the stellar mass also increases, meaning that gas inflows to each galaxy keeps the gas content from depletion. The left panel on the second row shows the galactic star formation rate (SFR). The SFR shows the largest peaks ($\gtrsim 100\ \mathrm{M_\odot/yr}$) for systems A, B, D, E and G. These (apart from E) are also the systems that are part of galaxy mergers and the evolution of their SMBH binaries are studied in Section \ref{sect:cosmo_bh}. The evolution of the SMBH mass accretion rates and the SMBH masses are shown in the right panels of rows two and three, respectively. The initial accretion rates at the moment of BH seeding are low (below $10^{-3}\,\mathrm{M_\odot/yr}$) and as was already seen in Figure \ref{fig:bh-stellar-relation}, the SMBHs initially do not grow in mass by a significant amount.

The left panel of the third row shows the evolution of the effective radius of the stellar component and the bottom row shows the gas fraction calculated from within either the three-dimensional stellar half-mass radius $r_{1/2,\star}$ (left panel) or $r_\mathrm{gal}$ (right panel). As the phase of compaction begins, both the gas fractions within the half-mass radius and $r_\mathrm{gal}$ sharply decrease. For the gas fraction calculated within the central half-mass radius, the values drop below 10\% for all systems, with most (all except systems C, G and H) having central gas fractions varying between 0.1\% and 3.0\%. In contrast to the central regions, the galaxy does not get depleted from gas. The total gas fraction of each galaxy begins to decrease as the compaction phase starts, but the gas fractions always remain between 20\% and 50\%. Such an evolution of the gas fraction also explains the size evolution of the galaxies. The compaction is caused by a centrally concentrated star formation, ending with the central region being depleted from gas. The outer regions of each galaxy still include a large fraction of gas, leading to subsequent star formation occurring off-centre and thus increasing the effective radius of the stellar component. In addition, constant gas inflows to each galaxy keeps the gas supply from depleting as the star formation continues.
 
\subsection{Galaxy metallicities}

The oxygen abundance is often used as a tracer for the total metallicity of a galaxy in observations (e.g. \citealt{1999ApJ...511..118K,Tremonti2004}). In Figure \ref{fig:metallicity} we show the stellar metallicity as measured by $12+\log(\mathrm{O/H})$ as a function of the stellar mass for all the simulated systems with \ketju{} integrated SMBHs, extracted from the zoom-K+DF simulation. We also overplot JWST observations of 135 galaxies in the redshift range $4<z<10$, which were used in \cite{Nakajima2023} for the mass--metallicity relation calculation. 

The star formation in the simulated compact galaxies result in a metallicity evolution that agrees well with the observations throughout their evolution. Until a stellar mass of $\sim 10^9\,\mathrm{M}_\odot$ is reached, the metallicity monotonically increases with stellar mass, with the metallicity reaching values as large as $12+\log(\mathrm{O/H})\sim8.5$. Only a few data points exist for the data set of \citet{Nakajima2023} at masses larger than a few times $10^9\, \mathrm{M}_\odot$. At such large stellar masses, the metallicities of our simulated galaxies reach a plateau and do not keep increasing with stellar mass. This plateau continues to our largest stellar mass of $\sim 10^{11}\ \mathrm{M}_\odot$. Although the galaxies in \cite{Nakajima2023} do not reach the masses where the plateau occurs, observations from lower redshift galaxies (e.g. \citealt{Tremonti2004, Sanchez2015, Curti2020}) show a similar trend. 

\begin{figure*}
    \centering

    \begin{minipage}[b]{0.6\textwidth}
        \begin{minipage}[b]{0.3\textwidth}
            \centering
            \includegraphics[width=\linewidth]{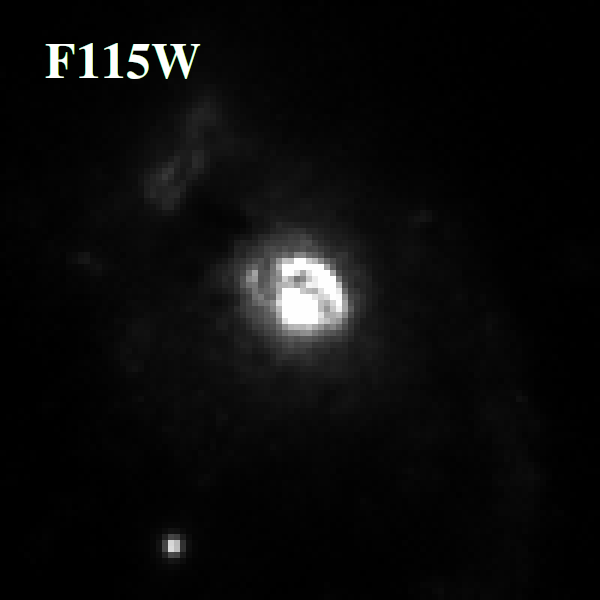}
        \end{minipage}
        \begin{minipage}[b]{0.3\textwidth}
            \centering
            \includegraphics[width=\linewidth]{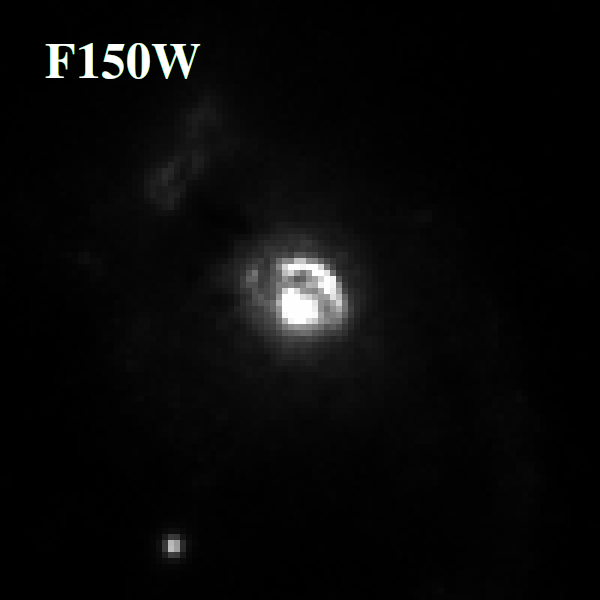}
        \end{minipage}
        \begin{minipage}[b]{0.3\textwidth}
            \centering
            \includegraphics[width=\linewidth]{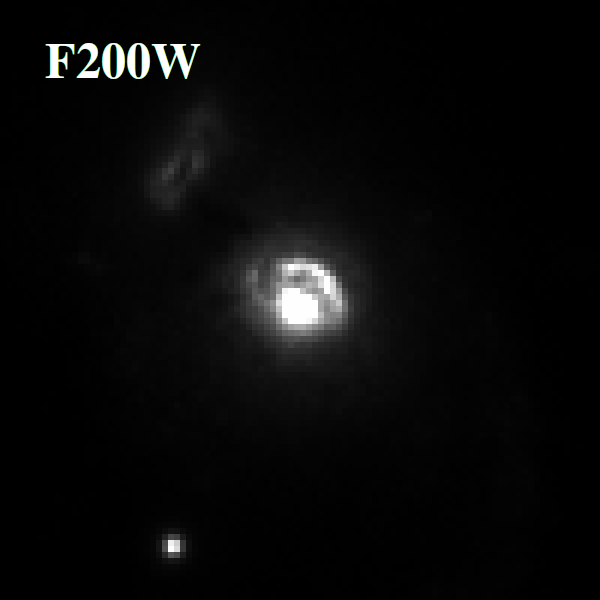}
        \end{minipage}

        \begin{minipage}[t]{0.3\textwidth}
            \centering
            \includegraphics[width=\linewidth]{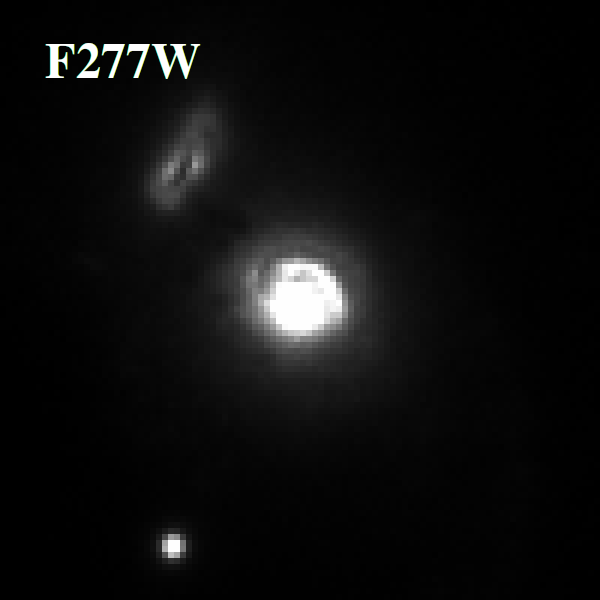}
        \end{minipage}
        \begin{minipage}[t]{0.3\textwidth}
            \centering
            \includegraphics[width=\linewidth]{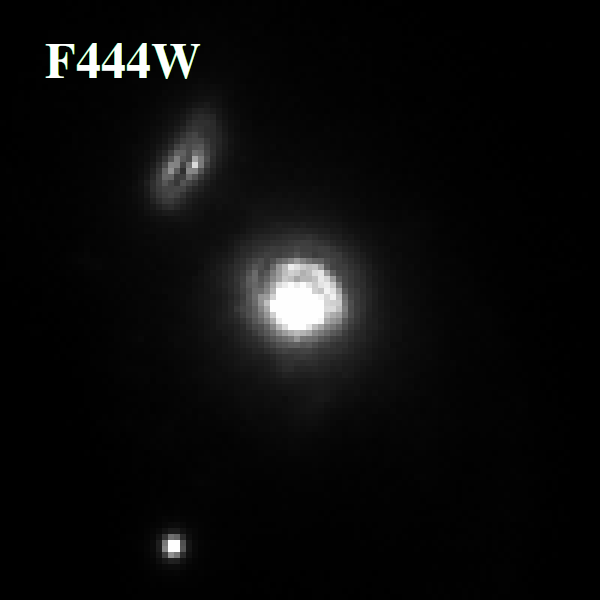}
        \end{minipage}
        \begin{minipage}[t]{0.3\textwidth}
            \centering
            \includegraphics[width=\linewidth]{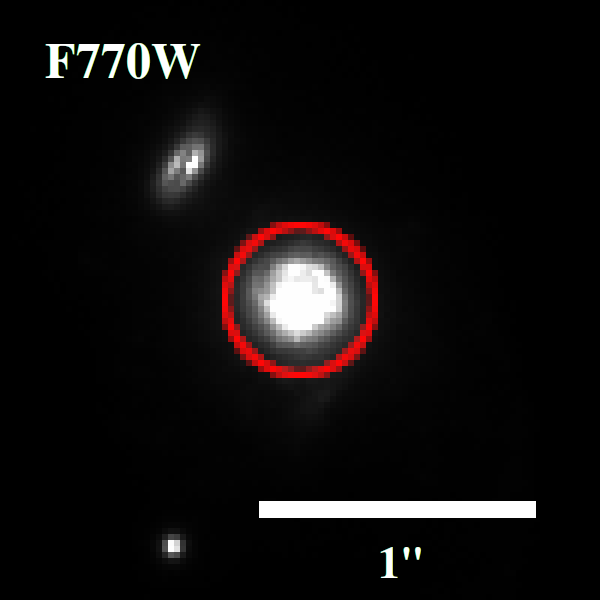}
        \end{minipage}
    \end{minipage}
    \begin{minipage}[b]{0.25\textwidth}
        \centering
        \includegraphics[width=\linewidth]{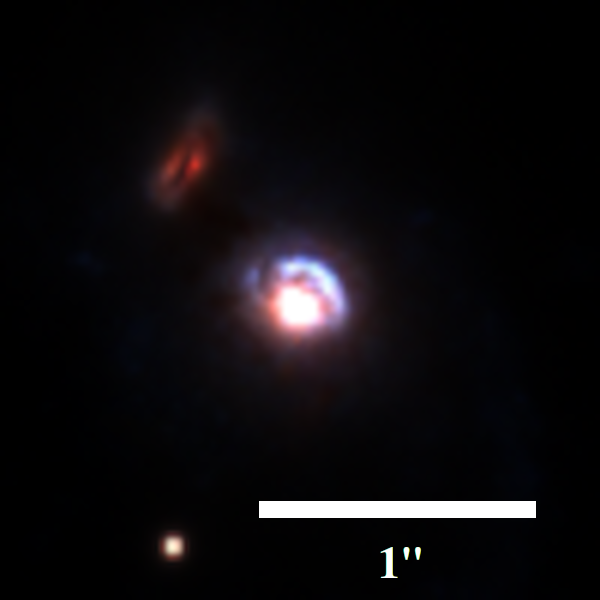} 
    \end{minipage}

    \begin{minipage}[t]{\textwidth}
        \centering
        \includegraphics[width=\linewidth]{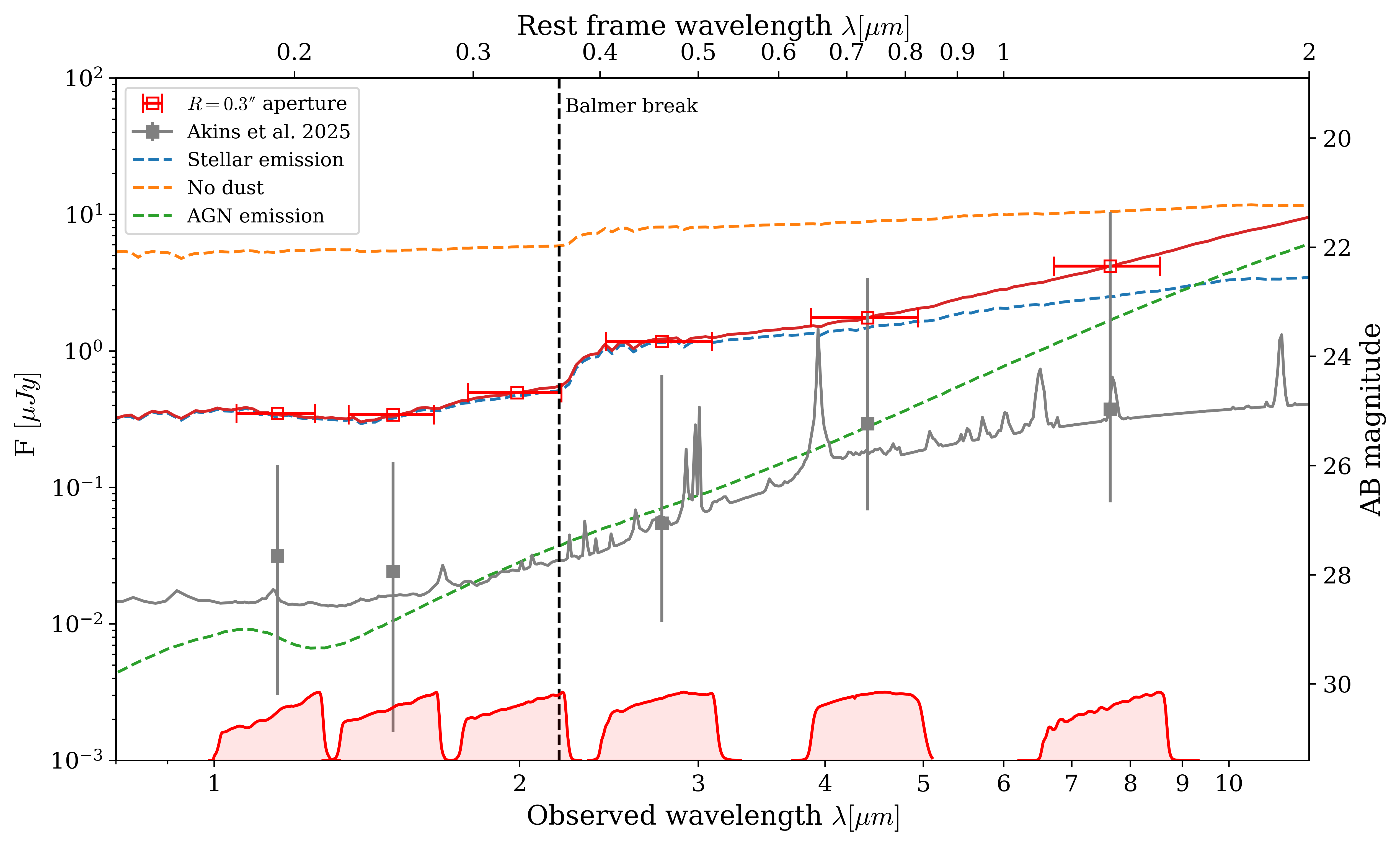}
    \end{minipage}

    \caption{Mock SKIRT images and SED of galaxy A at a redshift of $z=5$. The upper panels show broadband images in six different JWST bands, NIRCam F115W, F150W, F200W, F277W and F444W, and MIRI F770W, as well as an RGB image of three bands, F115W/F200W/F444W. The red circle in the MIRI F770W image indicates the aperture size $R_\mathrm{aperture} = 0.3''$ used for the photometry, while the white bars in the F770W and RGB colour images correspond to one arcsecond. The SED was computed both directly with SKIRT (red solid line) and using aperture photometry centered on the JWST broadbands shown in the upper panels (red squares). The error bars in the squares indicate the effective width of each JWST band, corresponding to the red transmission curves shown at the bottom of the figure. The dashed orange line corresponds to the SED in the case that there is no dust in the system. The blue and green dashed lines correspond to emission from stars and AGN only, respectively. The grey squares are median values from \citet{2025ApJ...991...37A} for observed galaxies within a redshift range of $z=5\pm 0.5$, with the vertical lines showing the full range of values around the median. The grey SED is the stacked SED from \citet{2025ApJ...991...37A} for all objects, moved to the corresponding observed frame. The vertical dashed line shows the location of the Balmer break in the observed frame.}
    \label{fig:SKIRT_galaxyA}
\end{figure*}

\begin{figure*}
    \centering
    \begin{minipage}[b]{0.6\textwidth}
        \begin{minipage}[b]{0.3\textwidth}
            \centering
            \includegraphics[width=\linewidth]{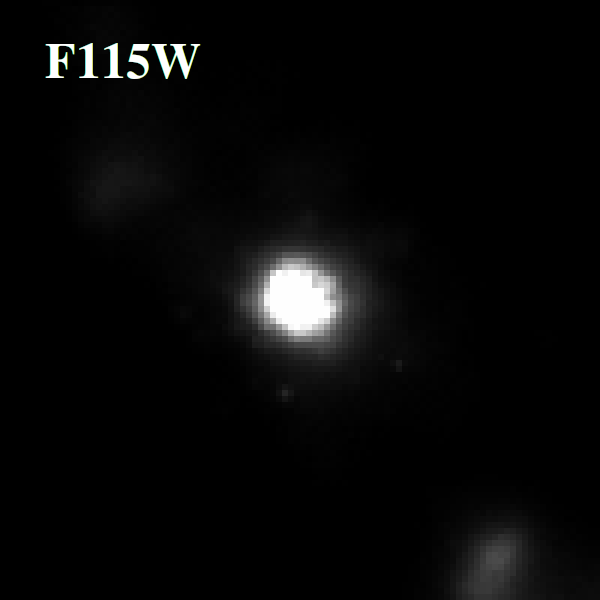}
        \end{minipage}
        \begin{minipage}[b]{0.3\textwidth}
            \centering
            \includegraphics[width=\linewidth]{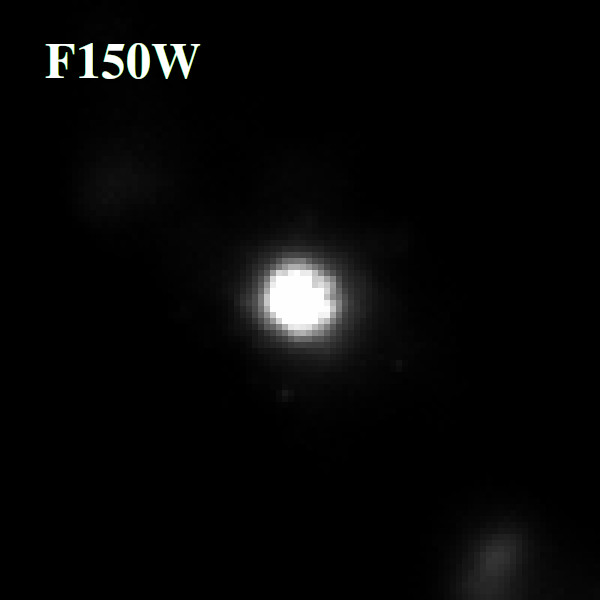}
        \end{minipage}
        \begin{minipage}[b]{0.3\textwidth}
            \centering
            \includegraphics[width=\linewidth]{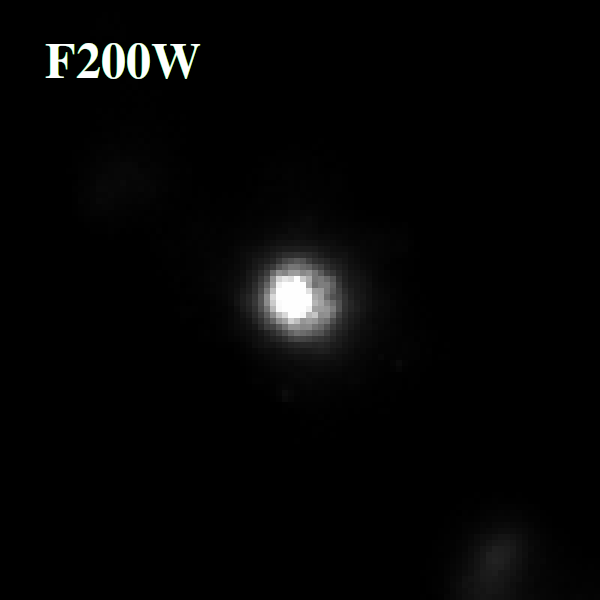}
        \end{minipage}

        \begin{minipage}[t]{0.3\textwidth}
            \centering
            \includegraphics[width=\linewidth]{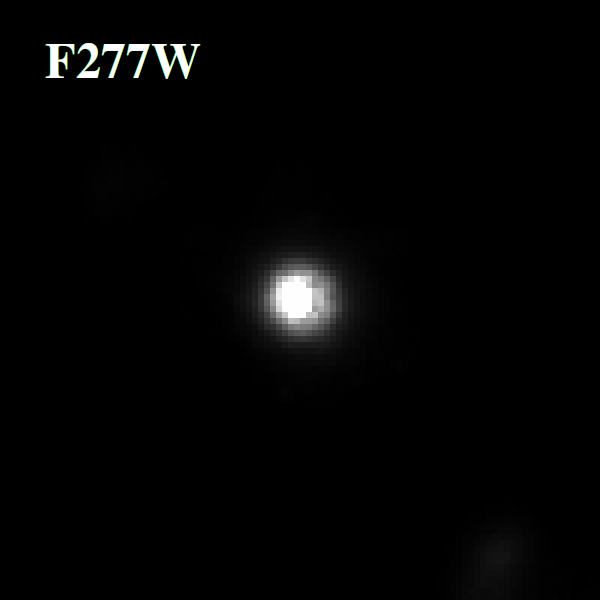}
        \end{minipage}
        \begin{minipage}[t]{0.3\textwidth}
            \centering
            \includegraphics[width=\linewidth]{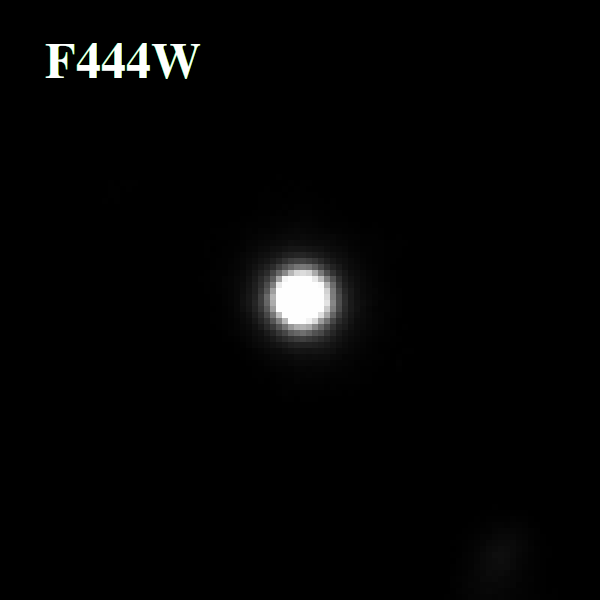}
        \end{minipage}
        \begin{minipage}[t]{0.3\textwidth}
            \centering
            \includegraphics[width=\linewidth]{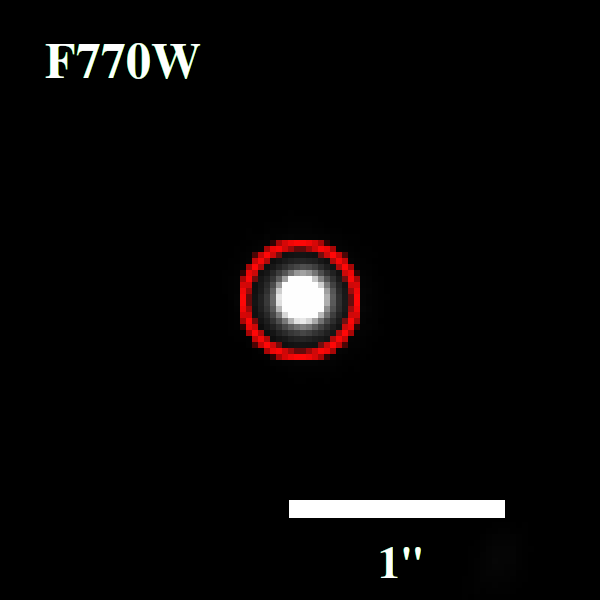}
        \end{minipage}
    \end{minipage}
    \begin{minipage}[b]{0.25\textwidth}
        \centering
        \includegraphics[width=\linewidth]{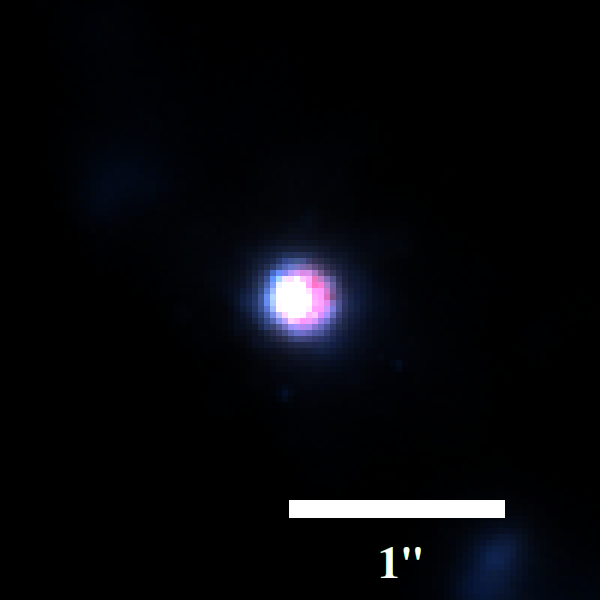} 
    \end{minipage}

    \begin{minipage}[t]{\textwidth}
        \centering
        \includegraphics[width=\linewidth]{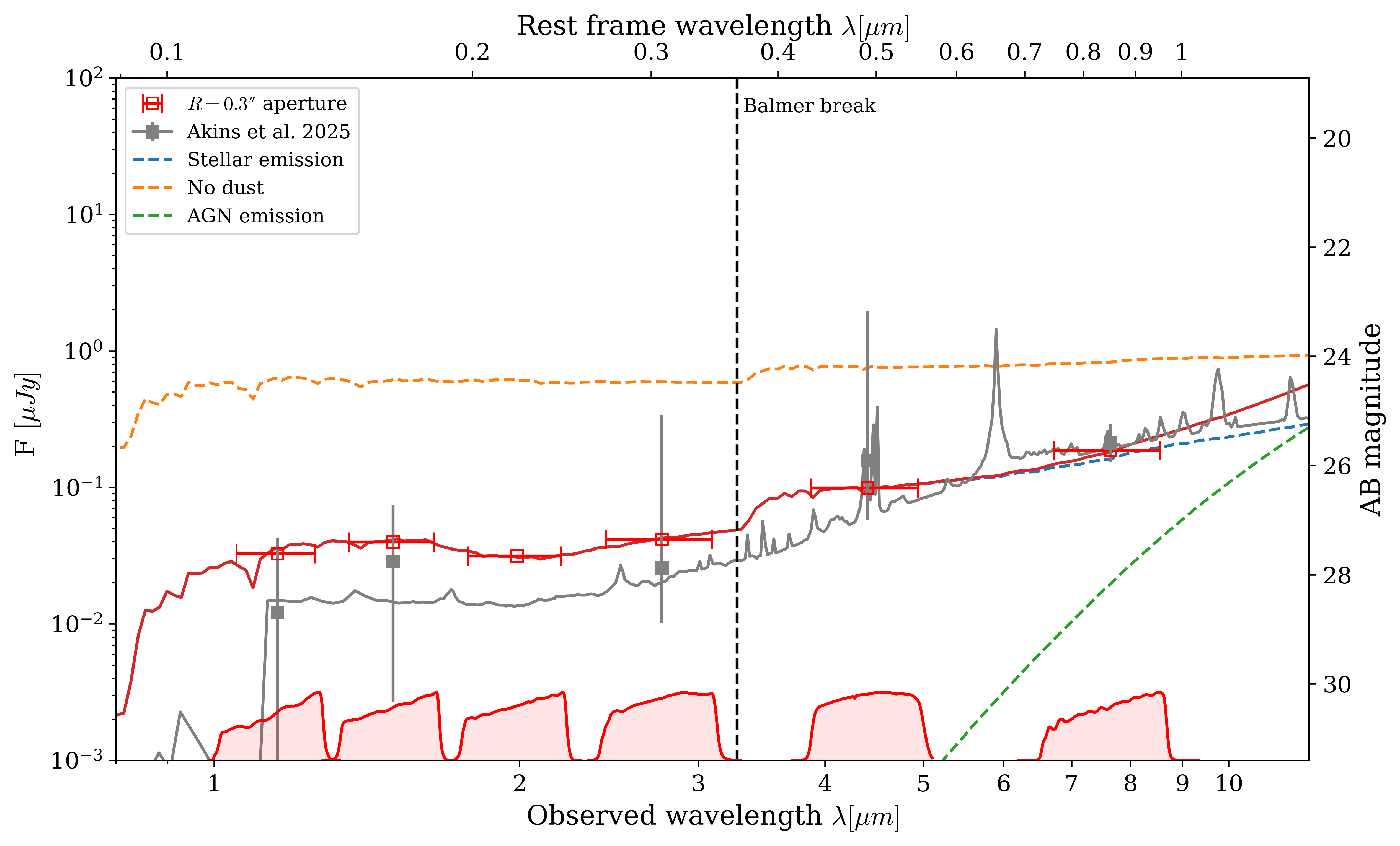}
    \end{minipage}
    \caption{Mock SKIRT images and SED of galaxy B at redshift $z=7.986$. As in Fig. \ref{fig:SKIRT_galaxyA} the upper panels show broadband JWST mock images, with the observed full SED shown in the main panel (red line, with red squares). Again, the blue and green dashed lines correspond to emission from stars and AGN only, respectively, with the orange dashed line showing the no dust case for the full SED. The grey squares are median values from \citet{2025ApJ...991...37A} for observed galaxies within a redshift range of $z=8\pm 0.5$, with the vertical lines showing the full range of values around the median. Finally, the grey SED is the stacked SED from \citet{2025ApJ...991...37A} for all objects, moved to the corresponding observed frame. The vertical dashed line shows the location of the Balmer break in the observed frame.}
    \label{fig:SKIRT_galaxyB}
\end{figure*}

\subsection{Mock JWST images and SEDs}

In order to produce mock JWST images and spectral energy distributions (SEDs), we used the radiative transfer model \textsc{SKIRT} \citep{Camps&Baes:2020}. \textsc{SKIRT} uses a Monte Carlo technique to trace the emission of light from different sources, its absorption, and subsequent re-emission by gas and dust. 

As representative examples, we chose to model galaxy A at the end of the simulation, at redshift $z=5$ (see Figure \ref{fig:SKIRT_galaxyA}), and galaxy B at a redshift of $z=7.986$ (see Figure \ref{fig:SKIRT_galaxyB}), when the galaxy was undergoing compaction. For each galaxy, we used the stellar particles from the simulation and a point source representing the AGN as sources of emission. The emission from stars was modelled within \textsc{SKIRT} by computing a SED for each stellar particles from their metallicity, age, and mass using the population synthesis model by Bruzual \& Charlot \citep{Bruzual2003}, assuming a Salpeter \citep{1955ApJ...121..161S} initial mass function. For the emission from the AGNs, we used a typical quasar SED template provided in \textsc{SKIRT}, based on \citet{Stalevski:2012} and \citet{Schartmann:2005}, with a Netzer emission profile \citep{Netzer:1987}. For a Netzer AGN profile, the emission originates from a geometrically thin, but optically thick disk, with an angle-dependent axisymmetric emission. The luminosity is brightest when viewed face-on, and diminishes as a function of the inclination angle $\theta$. To avoid either extreme (face-on, $\theta=0^{\circ}$, and edge-on, $\theta=90^{\circ}$), we set the inclination angle to be $\theta=30^{\circ}$ from the face-on axis. Finally, for normalising the AGN SEDs, we computed the bolometric luminosity of each AGN based on their instantaneous mass accretion rates at the corresponding snapshots, resulting in $L_\mathrm{A,bol}=5.99\times10^{45}\, \mathrm{erg / s} = 1.56\times10^{12}\, \mathrm{L}_{\odot}$ and $L_\mathrm{B,bol}=9.72\times10^{44}\, \mathrm{erg / s} = 2.54\times10^{11}\, \mathrm{L}_{\odot}$ for the AGNs in galaxies A and B, respectively. 

For the interstellar medium, we used directly the simulation SPH particles. The dust mass contained within each particle was calculated as a fraction of the metallicity times the total mass of the gas particle. Following \citet{Dwek:1998} we assumed a dust-to-metallicity fraction of 0.3, which resulted typically in a dust content below 1\% of the gas mass. We also set an upper limit to the temperature at which dust can exist, by assuming that any gas particle with a temperature above $T_{\rm gas}>10^{4} \ \rm K$ contains no dust. For modelling the dust, we used the \textsc{THEMIS} \citep{Jones2017:THEMIS} dust model provided within \textsc{SKIRT}.

\textsc{SKIRT} produced as outputs for each galaxy a SED over all wavelengths and a FITS datacube, containing 2D images of the galaxy for a set of wavelength bins within a given range. The field-of-view of the images is 15 kpc in the local frame of the galaxies, corresponding to $2.34''$ at redshift $z=5$ (galaxy A) and $3.05''$ at redshift $z=7.986$ (galaxy B). To obtain JWST-like images, the datacubes were convolved over the NIRCam F115W, F150W, F200W, F277W, and F444W, and MIRI F770W bands, resulting in six broadband images (upper panels in Figures \ref{fig:SKIRT_galaxyA} and \ref{fig:SKIRT_galaxyB}). In addition, we produced an RGB image combining three bands, F115W, F200W, and F444W. 

The broadband images were used to measure the flux and magnitudes using a circular aperture of radius $R_\mathrm{aperture}=0.3''$ (red circles in the MIRI F770W images in Figures \ref{fig:SKIRT_galaxyA} and \ref{fig:SKIRT_galaxyB}). We used the same aperture for all broadband images of both the A and B galaxies. The obtained fluxes are shown as red squares in the main panel of Figures \ref{fig:SKIRT_galaxyA} and \ref{fig:SKIRT_galaxyB}, with the horizontal error bars representing the widths of the transmission curves of each broadband. The transmission curves are also plotted separately at the bottom of the figures. The solid red line represents the total SED within the aperture $R_\mathrm{aperture}$, while the dashed blue and green lines show the individual contributions of the stellar and AGN emission, respectively. To calculate the emission from stars, we rerun \textsc{SKIRT} without the AGN component, i.e. by including only stars, dust and gas. The AGN contribution was obtained correspondingly, by rerunning \textsc{SKIRT} without the inclusion of the stellar component. To assess the impact of the dust component for the obscuration and re-emission of flux, we also rerun \textsc{SKIRT} for the full stellar+AGN model without dust (dashed red line in the main panel of Figures \ref{fig:SKIRT_galaxyA} and \ref{fig:SKIRT_galaxyB}). 

We compared our SEDs to observations of LRDs presented in \citet{2025ApJ...991...37A}, shown in grey in the main panels of Figures \ref{fig:SKIRT_galaxyA} and \ref{fig:SKIRT_galaxyB}. For each figure, we selected galaxies within a redshift range of $\Delta z=\pm0.5$ of our simulated galaxy. The squares are the median fluxes within the broadbands NIRCam F115W, F150W, F277W, and F444W, and MIRI F770W, with the vertical lines indicating the range from minimum to maximum values in the observed sample. The grey solid line corresponds to the stacked SED from all observations \citep[Figure 13 in][]{2025ApJ...991...37A}, here shifted to the corresponding observed frame.

Studying Figures \ref{fig:SKIRT_galaxyA} and \ref{fig:SKIRT_galaxyB}, we see that the SEDs are in both cases dominated by the stellar component, with the AGN emission contributing substantially only at relatively long wavelengths of $\lambda\gtrsim 5 \ \mu \rm m$. For system A, the SMBH mass is $m_\bullet=3.2\times10^8\,\mathrm{M}_\odot$ and the accretion rate is $\dot{m}_\bullet=1.1\,\mathrm{M}_\odot/\mathrm{yr}$, leading to Eddington fraction of $f_\mathrm{Edd}=\dot{m}_\bullet/\dot{m}_\mathrm{Edd}=0.15$. Interestingly, the SMBH of galaxy B at $z=7.986$ is accreting at the Eddington limit ($f_\mathrm{Edd}=0.998$, $\dot{m}_\bullet=0.15\,\mathrm{M}_\odot/\mathrm{yr}$), but has a mass that is almost two orders of magnitude smaller ($m_\bullet=7.7\times10^6\,\mathrm{M}_\odot$) than system A, resulting also in a significantly lower AGN luminosity. However, we stress that different choices for the AGN modelling, including changing the inclination angle in the employed Netzer model could considerably affect the resulting AGN flux. In addition, we are also showing here a relatively narrow wavelength window in which the stellar emission in general dominates over the AGN. We also find that the SED of galaxy A at $z=5$ lies somewhat above the stacked SED (grey lines) from \citet{2025ApJ...991...37A}, which is not too unexpected, given that galaxy A is the most massive object in our simulation volume. The SED of galaxy B at $z=7.986$, on the other hand, is in general good agreement with the stacked SED both for the fluxes and the general SED shape. However, a more detailed comparison would require an analysis of individual emission lines (e.g. \citealt{2025MNRAS.tmp.1995D,2025arXiv250311752D}), which is not included in our SKIRT SED model.
Nevertheless, this indicates that our simulated galaxies, in addition to their masses and sizes, also have SED properties that are in general agreement with the observed LRD population.   

\section{Black holes in high redshift galaxies}
\label{sect:cosmo_bh}

In this section we study the evolution of the \ketju{} SMBH binaries that were resolved in the zoom-in simulation. We focus on the merger timescales and the evolution of the properties of the gravitationally bound binaries. We also show that by using \ketju{} together with a phenomenological model for the final few orbits of a SMBH binary, 
the GW emission can be modelled through nanohertz frequencies (which PTAs target) to micro- and milliherz frequencies that LISA will observe. Before focusing on SMBH binaries, we first validate that the combination of \ketju{} and the dynamical friction subgrid model is capable of keeping SMBHs at the centres of their host galaxies, and that the mass limit for \ketju{} integration $m_\mathrm{Ketju}$ is sufficient for isolated SMBHs also in a cosmological environment.

\subsection{Black hole displacement}
\label{subsect:cosmo_bh_dyn}

\begin{figure}
    \centering
    \includegraphics[width=\linewidth]{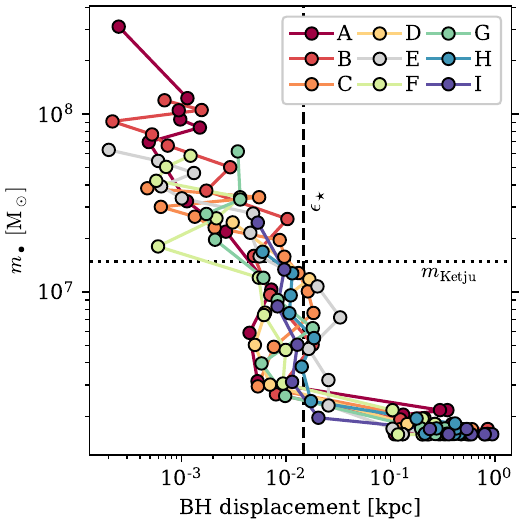}
    \caption{The evolution of mass and separation from the centre of the host galaxy from the simulation zoom-K+DF for each SMBH which reaches the \ketju{} integration mass limit, shown as a horizontal dashed line. The vertical dashed line shows the physical stellar softening length $\epsilon_\star$. The results are from snapshots with an interval of $\sim 60\ \mathrm{Myr}$. Three distinct stages of evolution are visible: seed mass SMBHs orbit around the centre and eventually sink due to dynamical friction, after which the subgrid model keeps them at a separation of $\sim\epsilon_\star$. When a SMBH reaches the mass limit $m_\mathrm{Ketju}$, the separation from the centre of the host galaxy decreases even further and can reach values below 1 pc.}
    \label{fig:displacements}
\end{figure}

In order to ascertain that the chosen mass limit $m_\mathrm{Ketju}$ is not too low in a cosmological simulation for the SMBHs to remain at the centre when \ketju{} is enabled, we show in Figure \ref{fig:displacements} the mass and displacement for each of the nine SMBHs which satisfies the criteria $m_\bullet\geq m_\mathrm{Ketju}$. The interval between data points is $\sim 60\ \mathrm{Myr}$. In addition, we show in \autoref{table:properties} the redshifts at which a SMBH was seeded and when it reached the \ketju{} integration mass limit, together with the properties of the their host galaxies.

In Figure \ref{fig:displacements}, the vertical axis shows the SMBH mass and the horizontal axis the distance from the centre of the host galaxy, calculated with the shrinking sphere method \citep{shrinking_sphere_Power03} using the stellar particles. Again, the results are from the simulation zoom-K+DF and are shown for the systems A-I which have a SMBH reaching mass $m_\mathrm{Ketju}$ during the simulation. 
We find that the SMBHs orbit around the centres of their host galaxies when the SMBHs are still around the seed mass of $10^6\ \mathrm{M}_\odot$. At the moment of seeding, the host galaxies typically have small masses (the baryonic masses are typically below $10^8\ \mathrm{M}_\odot$, see Figure \ref{fig:multipanel}) resulting in  relatively shallow gravitational potentials. Seed mass SMBHs still remain typically within a radius of $1\,\mathrm{kpc}$ from the centre of the galaxy. The galaxies continue to grow in mass and deepen their central potentials and eventually dynamical friction sinks the seed mass SMBHs to the centres of their host galaxies. The SMBH growth starts as the seed mass SMBH reaches the centre (see Figure \ref{fig:bh-stellar-relation} for the stellar masses of galaxies at which the SMBH growth begins), and after this the SMBH remains in the central region
of their host galaxies.

\begin{figure*}
    \centering
    \includegraphics[width=\linewidth]{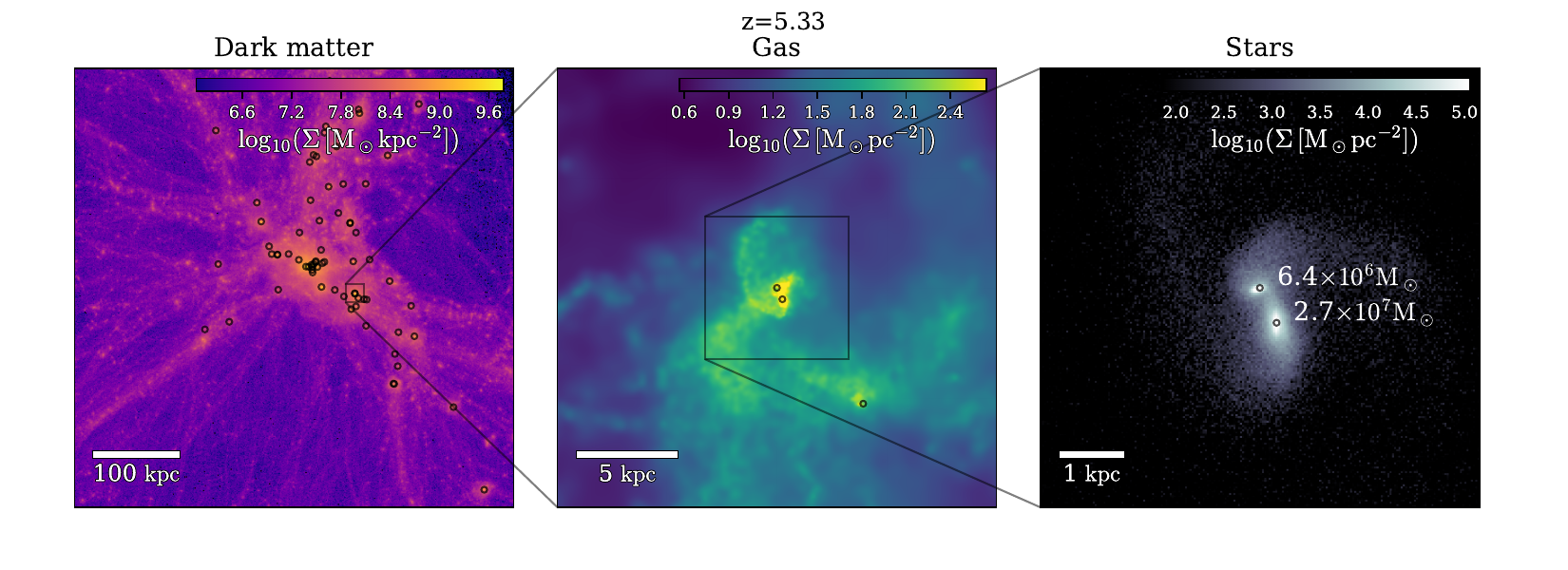}
    \caption{Snapshot from redshift $z=5.33$. Left: surface density profile of dark matter in a part of the zoom-in region. The black circles represent the positions of SMBHs. Middle: surface density of the gas particles from a region including an ongoing galaxy merger. Right: surface density of the stellar particles of the ongoing galaxy merger. The galaxy merger will result a SMBH binary (merger J$\rightarrow$G), with the coalescence of the SMBHs occurring 44 Myr after the shown snapshot. At $z=5.33$ the smaller SMBH has a mass of $m_\bullet=6.4\times 10^6\ \mathrm{M}_\odot$ while the more massive SMBH the mass is $m_\bullet=2.7\times 10^7\ \mathrm{M}_\odot$.
    }
    \label{fig:zoom-in}
\end{figure*}

From Figure \ref{fig:displacements} we also see that as the dynamics of a SMBH starts to be modelled with \ketju{}, there is a clear trend in the magnitude of the displacement. After the SMBHs have begun to grow, their separation from the centre is slightly below the stellar softening, similarly as was seen for the isolated systems in Section \ref{sect:Brownian}. As the \ketju{} mass limit is reached, displacements generally become of the order of $\sim 1 \ \rm pc$. This is also in agreement with what was seen in the isolated simulations in Figure \ref{fig:brownian_motion}: the Brownian motion with \ketju{} decreases as the mass of the SMBH is increased, while the Brownian motion keeps the SMBH roughly at the same radius regardless of its mass when the subgrid model is used. The fact that the displacement does not grow when switching the model for the SMBH dynamics indicates that the switching from the dynamical friction subgrid model to \ketju{} integration works as intended in a cosmological setting.

\subsection{Black hole binaries}

\begin{figure*}
    \centering
    \includegraphics[width=0.8\linewidth]{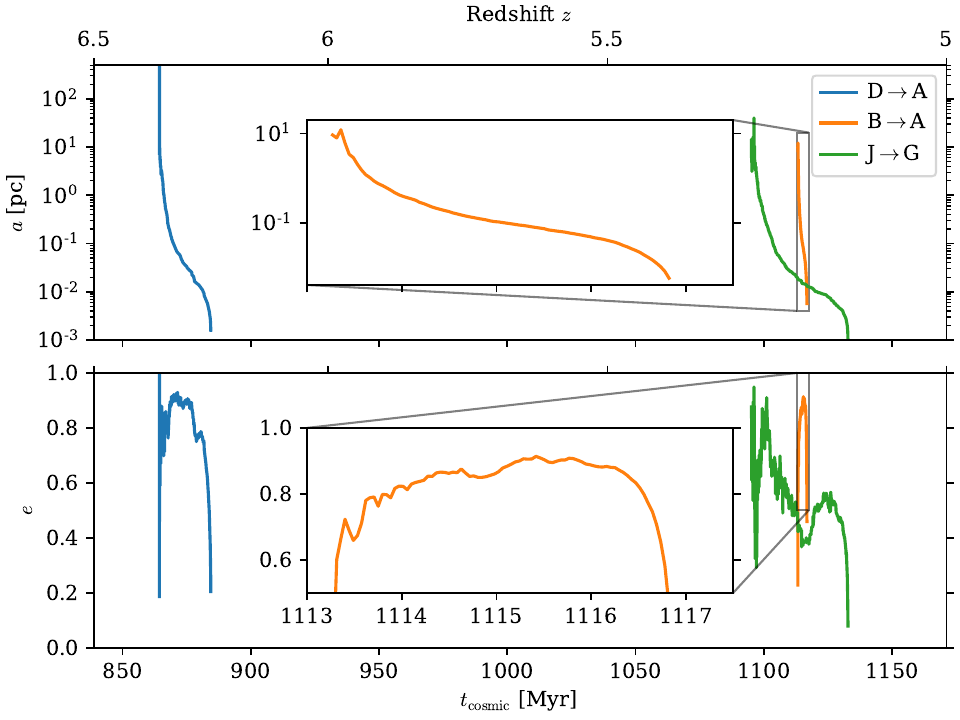}
    \caption{Binary parameters (top: semimajor axis and bottom: eccentricity) of the \ketju{} integrated SMBH binaries in the cosmological zoom-in simulation. Zoomed panels are included for the system B$\rightarrow$A, which reaches coalescence just $\sim 3.5\,\mathrm{Myr}$ after becoming bound.}
    \label{fig:zoom-in_binary_params}
\end{figure*}

In the left panel of Figure \ref{fig:zoom-in} we show the projected dark matter density of a central part of the zoom-in region, with the locations of the SMBHs shown as circles. The middle panel zooms into a region with an ongoing galaxy merger and shows the surface density of the gas component. A further zoom is shown in the right panel and shows the stellar surface density of the ongoing galaxy merger. The SMBH binary which will be formed as a result of the galaxy merger (binary J$\rightarrow$G in forthcoming figures) is one of the \ketju{} integrated binaries in the zoom-K+DF simulation. 

In Figure \ref{fig:zoom-in_binary_params} we show the evolution of the semimajor axis (top panel) and 
the eccentricity (bottom panel) of the three \ketju{} integrated binaries. 
Each line starts from the moment after which each binary remains gravitationally bound. The mergers are labelled as D$\rightarrow$A, B$\rightarrow$A and J$\rightarrow$G,
where the letter before the arrow states that the system merged away and the letter after the arrow is the system which is left after the merger of the two SMBHs. Binaries D$\rightarrow$A and B$\rightarrow$A possess large eccentricities, reaching values in excess of $e \gtrsim0.9$. During the GW-dominated phase of the binary evolution, the eccentricity drops. The first \ketju{} integrated binary (D$\rightarrow$A, $m_1=4.0\times 10^7\ \mathrm{M}_\odot$ and $m_2=2.8\times 10^7\ \mathrm{M}_\odot$ at the moment of merger) coalesces at a redshift of $z=6.24$, and the other two near the target redshift ($z=5$) of the simulation with binary B$\rightarrow$A ($m_1=1.5\times 10^8\ \mathrm{M}_\odot$, $m_2=1.5\times 10^8\ \mathrm{M}_\odot$) coalescing at $z=5.20$ and J$\rightarrow$G ($m_1=3.4\times 10^7\ \mathrm{M}_\odot$, $m_2=1.4\times 10^7\ \mathrm{M}_\odot$) at $z=5.16$.

\begin{figure}
    \centering
    \includegraphics[width=\linewidth]{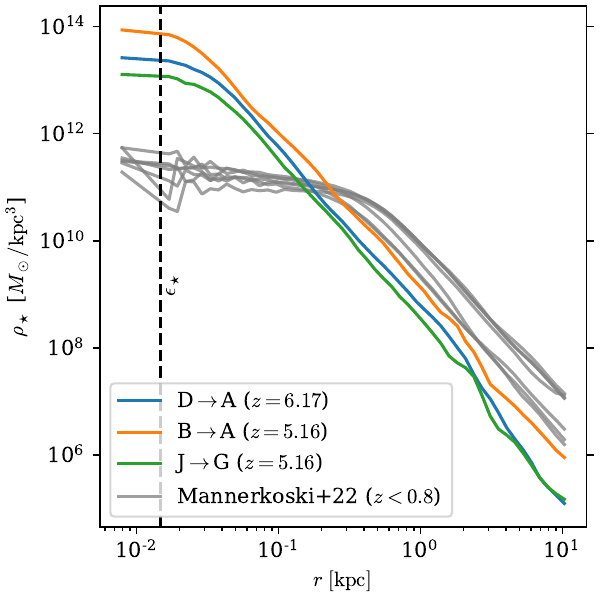}
    \caption{The three-dimensional stellar densities of galaxies, measured from the first snapshot after a \ketju{} integrated SMBH merger has occurred in the galaxy. D$\rightarrow$A, B$\rightarrow$A and J$\rightarrow$G show galaxies from the zoom-K+DF simulation performed in this study, the dashed line is the used stellar softening length $\epsilon_\star$ and the grey lines are galaxies from \citet{Mannerkoski2022}. In the central regions, the differences in stellar densities can reach up to two orders of magnitude. 
    }
    \label{fig:3d_densities}
\end{figure}

All three \ketju{} integrated binaries coalesce in less than $50\,\mathrm{Myr}$ after becoming gravitationally bound. All galaxies which contain \ketju{} integrated SMBH mergers have nearly zero central gas fractions (see Figure \ref{fig:multipanel}) and the main driver for the rapid mergers is the high central stellar densities. In Figure \ref{fig:3d_densities} we show the three-dimensional stellar density profiles of the galaxies where the mergers occur. The densities are calculated from the first snapshots after the SMBHs have merged. As a comparison, we also include densities of galaxies from the \ketju{} integrated mergers from a cosmological zoom-in simulation presented in \cite{Mannerkoski2022}. The coalescence for most of these binaries took several hundred Myr with mergers taking place at relatively low redshifts of $z\sim 0.2-0.7$. Compared to the $z\lesssim 1$ galaxies from \cite{Mannerkoski2022}, the high redshift galaxy merger remnants in this study reach roughly two orders of magnitude higher stellar densities in the central $100\,\mathrm{pc}$. Such a large difference in the stellar densities, explains the rapid BH mergers, since larger stellar densities increase the rate of binary hardening \citep{Quinlan96, Liao_rabbits_I_2024, Liao_rabbits_II_2024}. 

\begin{figure*}
    \centering
    \includegraphics[width=\linewidth]{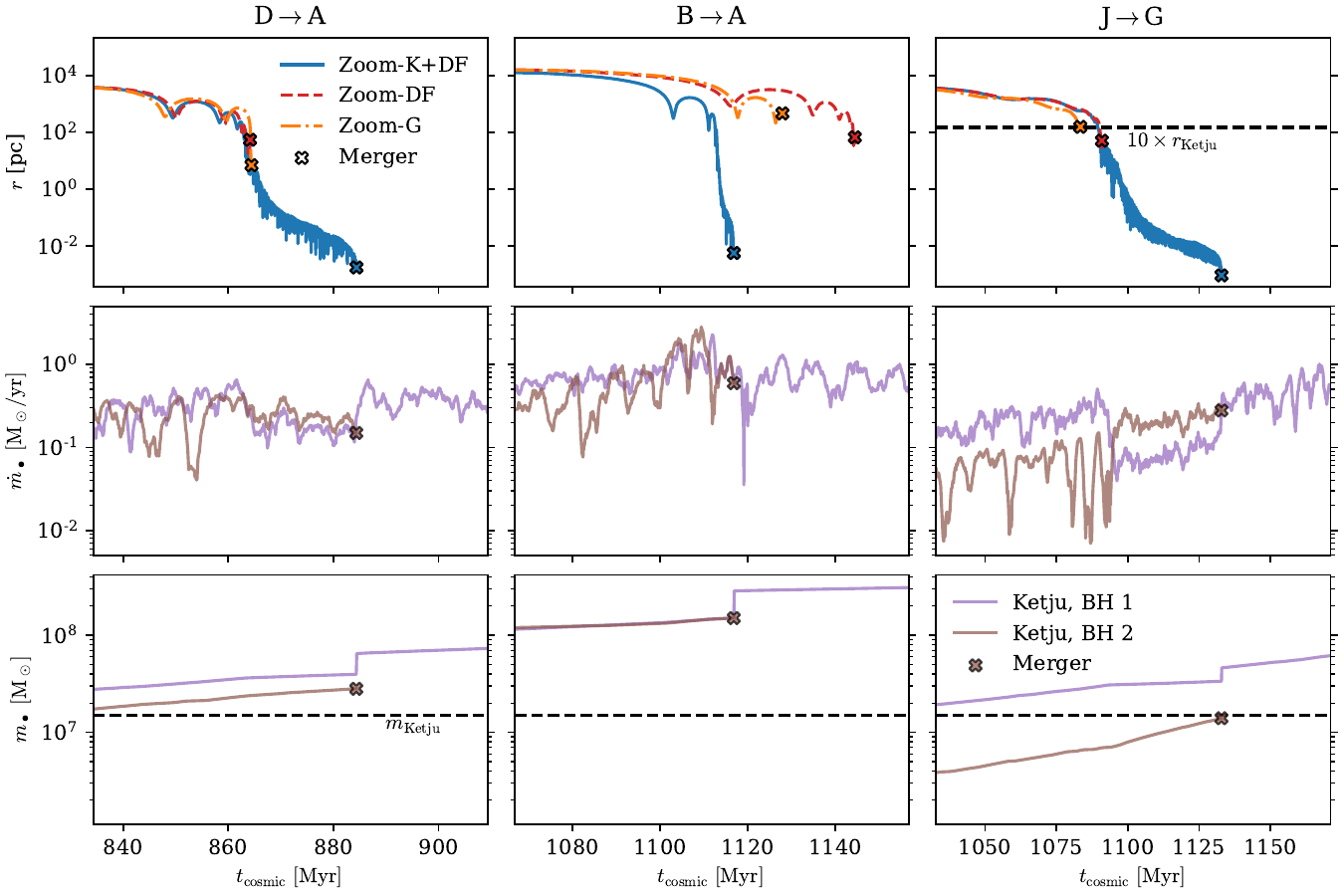}
    \caption{The evolution of the separation between the SMBHs (top row), the SMBH mass accretion rate (averaged over $1\,\mathrm{Myr}$, middle row) and the masses of the SMBHs (bottom row)  
    for binaries D$\rightarrow$A (left column), B$\rightarrow$A (middle column) and J$\rightarrow$G (right column) as a function of time. The solid lines show results from the simulation zoom-K+DF, while the dashed line shows results from the run zoom-DF and the dotted-dashed line is from the run zoom-G, with the latter two simulations only shown in the top row. For the merger J$\rightarrow$G, the smaller SMBH is below the \ketju{} mass limit and thus modelled with the subgrid model until it comes within $10\times r_\mathrm{Ketju}$ from the larger SMBH. This separation is shown as a black dashed line in the first row for this binary. 
    }
    \label{fig:binary_properties}
\end{figure*}

In Figure \ref{fig:binary_properties} we show the SMBH separations from all three simulations (top panel), the SMBH mass accretion rates (middle panel) and the SMBH masses (bottom panel). The bottom two rows show results only from the zoom-K+DF simulation. For the evolution of the separation for each SMBH binary, the results from the other two simulations, zoom-DF (dashed lines) and zoom-G (dashed-dotted lines), are also shown. Focusing on mergers D$\rightarrow$A and J$\rightarrow$G, we see that the systems simulated with \ketju{} and the dynamical friction subgrid model agree very well with each other when the separation is $r\gtrsim 100\,\mathrm{pc}$, after which SMBHs modelled with the subgrid model reach the artificial merger criteria. \ketju{} integrated binaries on the contrary reach sub-pc scale separations before their merger. The merger of system D$\rightarrow$A is delayed by 
$\sim 20\,\mathrm{Myr}$ and the merger of system J$\rightarrow$G by $\sim 35\,\mathrm{Myr}$ in the simulations including \ketju{}. 

Next, focusing on the merger B$\rightarrow$A, the two runs do not reach a separation of $r\sim 100\,\mathrm{pc}$ at the same time and the SMBH separations already differ during the first pericentre passage. This is due to the differences in the galactic-scale dynamics between the simulations. Although the simulations start from the same initial conditions, they are independent simulations, with different dynamical evolution, with these differences growing over time. The clear differences between the simulations for the region containing systems A and B are discussed in more detail in Appendix \ref{sect:run_diffs}. 

From the SMBH mass panels, we see that in binary J$\rightarrow$G the mass of the smaller SMBH is slightly below $m_\mathrm{Ketju}$ until the coalescence, meaning that its dynamics were modelled with the dynamical friction subgrid model until it reached a separation of $10\times r_\mathrm{Ketju}$ from the larger SMBH. Other binaries include only SMBHs which reach the $m_\mathrm{Ketju}$ mass limit before the separation of the two SMBHs is below $100\,\mathrm{pc}$. For binary D$\rightarrow$A, both of the SMBHs have reached a mass of a few times $10^7\,\mathrm{M}_\odot$ at the moment of merger. For B$\rightarrow$A, which reaches coalescence just $\sim 3.5\,\mathrm{Myr}$ after the binary becomes bound, both SMBHs have masses of $m_{\bullet}\sim 10^8\,\mathrm{M}_\odot$.

In the zoom-K+DF simulation the mass ratios of binaries D$\rightarrow$A and B$\rightarrow$A are relatively close to unity (at the moment of merger, $q\approx 0.713$ for D$\rightarrow$A and $q\approx 0.996$ for B$\rightarrow$A) but for binary J$\rightarrow$G the mass ratio is lower at $q\approx 0.367$ when the SMBHs coalesce. Lower values of $q$ lead to SMBH accretion being more preferential to the smaller SMBH (\autoref{eq:duffell}) and thus the effects of the SMBH binary accretion model are most visible for this binary. As the SMBH binary becomes gravitationally bound, the binary accretion model causes the accretion to favour the lower-mass SMBH and the accretion onto the more massive SMBH is reduced.

\subsection{Gravitational wave emission}

The final aspect we study for the SMBH binaries is their GW emission. Specifically, we calculate the characteristic strain from different times of the evolution of each binary, and see whether mergers of SMBHs in galaxies with sizes and masses matching JWST observations would be observable with the current PTAs and future LISA.

\begin{figure*}
    \centering
    \includegraphics[width=0.8\linewidth]{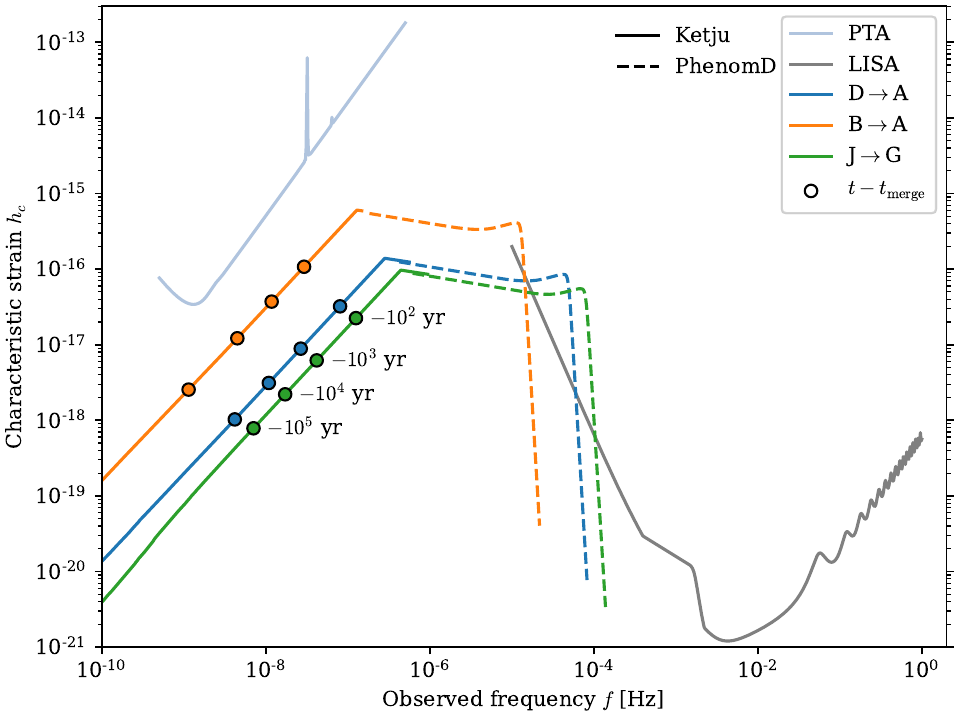}
    \caption{The characteristic strain as a function of the observed frequency for the three binaries, assuming an observation time of 30 yr. The dashed lines show the characteristic strain calculated with PhenomD assuming a dimensionless spin parameter of $\chi=0.8$ (the spin is set by hand to a value equal to the choice in \citealt{2020MNRAS.491.2301K}) and the solid lines show the evolution calculated from \ketju{} integrated binaries. The four dots for each binary gives the frequency and characteristic strain at four different times before the SMBH coalescence. These times, from left to right in the figure for each binary, are $10^5,\, 10^4,\, 10^3$ and $10^2\,\mathrm{yr}$. The grey solid line shows the LISA observation limit and the light blue solid line the PTA sensitivity curve. 
    }
    \label{fig:gw-strain}
\end{figure*}

In Figure \ref{fig:gw-strain}, we show the observed characteristic strain for the three binaries, assuming an observation time of $\tau=30\,\mathrm{yr}$. The solid lines show the evolution calculated directly from the \ketju{} simulation output. The details of how the characteristic strain is calculated are discussed in Appendix \ref{sect:gw_calc}. In addition, four dots in each line from left to right mark the GW strains and frequencies at times $10^5,\, 10^4,\, 10^3$ and $10^2\,\mathrm{yr}$ before the BH coalescence, respectively. Dashed lines are calculated using the PhenomD phenomenological code \citep{PhenomD-I, PhenomD-II}, which is able to model the characteristic strain of a binary during the final few orbits before coalescence. The characteristic strain is calculated using PhenomD assuming a dimensionless spin parameter of $\chi=0.8$, similarly as in \citet{2020MNRAS.491.2301K}. The figure also includes sensitivity curves for LISA (taken from PhenomD) and for PTAs. The PTA sensitivity curve is calculated with the Hasasia code \citep{Hazboun2019Hasasia}. For the PTA curve, we use 68 pulsars  with an observation time of 30 years, cadence of 3 weeks and timing noise of 200 nanoseconds. Although the observation time is longer than the current observations, the other parameters are taken to be similar to the current PTAs (e.g. \citealt{Nanograv15_obs_and_timing} and references therein, see also \citealt{2019ApJ...887...35M,Mannerkoski2022}).

Using the results from both \ketju{} and PhenomD, the complete evolution of GW emission by SMBH binaries can be calculated. The simulated binaries evolve through the frequency range observable with PTAs, however the calculated GW strains are far below the sensitivity level of current PTAs (see e.g. \citealt{2025arXiv250814742T} for a search of an individual binary source in the PTA data sets). There is a brief interval where the \ketju{} and PhenomD GW calculations overlap for mergers D$\rightarrow$A and J$\rightarrow$G. Here, the characteristic strains of the two models are in very good agreement, as the difference in the characteristic strains as calculated by \ketju{} and PhenomD only differ by less than $\lesssim5\%$. 

The characteristic strain of the final orbits before coalescence show that the mergers occur in the LISA frequency band, and are above the LISA sensitivity limit. Furthermore, the signal-to-noise ratios (SNRs) can be calculated by integrating over the frequency space of the LISA observation window, averaging over polarisation, inclination and sky location, via equation
\begin{equation}
    \mathrm{SNR} = \sqrt{\frac{32}{5}\int f^{-1}\frac{h_c^2(f)}{h_\mathrm{c,LISA}(f)} \mathrm{d}f},
\end{equation}
where $h_\mathrm{c,LISA}(f)$ is the characteristic strain observation limit for LISA at frequency $f$ \citep{2019CQGra..36j5011R}. The right hand side of the equation includes an extra factor of $\sqrt{2}$ due to LISA being a two-channel detector \citep{2019MNRAS.483.3108K}. For the merger, in which the SMBHs are the most massive (B$\rightarrow$A) the SNR is 2, but for mergers D$\rightarrow$A and J$\rightarrow$G the SNRs are 14 and 31, respectively. Therefore, coalescing SMBH binaries found in merging galaxies that are compact and observable with JWST would in principle be observable by LISA (assuming a standard SNR observability cut of 8 as in e.g. \citealt{2020MNRAS.491.2301K, 2023ApJ...955L..27N, 2025arXiv251009743S}). 

\section{Discussion}
\label{sect:discussion}

\subsection{Subgrid models of dynamical friction}

In Section \ref{sect:plummer_sphere}, we first studied the sinking of SMBHs in a Plummer potential using different dynamical friction models. We showed that the discrete dynamical friction subgrid model can lead to `double counting' for the effect of dynamical friction when the mass ratio between the SMBH and other particles is moderate. This is also discussed in the publication where the model was introduced \citep{Ma_discrete_subgrid_2023}, which focused on simulations where the mass ratio between the SMBH and other particles was set to a relatively low value of 10.

We decided to combine the \cite{Tremmel_subgrid_2015} dynamical friction subgrid mode with \ketju{}, because in this subgrid model the SMBHs were efficiently sinking to the centre of a system with mass ratios of $m_\bullet/m_\star=10$ and $100$, while also having Brownian motion equal to the stellar softening length. The subgrid model calculates the density used in the acceleration term (equation \ref{eq:a_df_tremmel}) using both stellar and dark matter particles, while \ketju{} applies regularised integration for SMBHs and nearby stellar particles. Since the evolution of binaries D$\rightarrow$A and J$\rightarrow$G agree well in terms of the evolution of the separation of the SMBHs (Figure \ref{fig:binary_properties}) between simulations zoom-K+DF and zoom-DF, the inclusion of dark matter in the dynamical friction calculation in these high redshift systems should have a negligible effect.


In contrast to the subgrid dynamical friction models, \ketju{} is able to model the interactions between SMBHs and stellar particles, including three-body scattering. For mergers of massive early-type galaxies with very massive SMBHs $(m_\bullet\gtrsim 10^{9} \ M_{\odot})$, this has been shown to cause cored density profiles in merger simulations \citep{rantala2018,2019ApJ...872L..17R,2025MNRAS.537.3421R}, which would not be seen if the SMBHs dynamics were modelled with a subgrid model. Resolving core formation in a cosmological simulation is challenging, and core formation was not seen in the cosmological zoom-in simulations of \citet{Mannerkoski2022} in which the majority of the SMBHs had masses in the range $m_{\bullet}\lesssim 10^{9} \ \rm M_{\odot}$. A large reason for this is also the relatively large stellar softening lengths compared to galaxy merger simulations, resulting in a softening-induced core before \ketju{} integrated mergers occurred. Although our simulation uses a smaller stellar softening length, the mergers in our simulation occur in systems with very high gas fractions and moderate SMBH masses $(m_{\bullet}\lesssim 10^{8} \ \rm M_{\odot})$. In addition, thermal AGN feedback is used for SMBHs, which is less effective at producing cores compared to kinetic AGN feedback \citep{Liao_binary_feedback_2023}, and core formation is thus not expected. In a future study we plan to continue simulations to lower redshifts and see whether we see noticeable differences in the stellar component of galaxies between simulations using \ketju{} and a dynamical friction subgrid model, and thus gain a better understanding of core formation in massive early-type galaxies.

\subsection{Galaxies and SMBHs at high redshift}
\label{sect:discussion_highz_gals}

In Section \ref{sect:gal_evo}, we first studied the evolution of the SMBHs on the $M_\star-m_\bullet$ relation. In our simulations the SMBHs follow the local relation of \cite{Kormendy2013} already at high redshifts and no overmassive SMBHs are produced. It is important to note that the SMBH mass estimates from JWST observations still include large uncertainties and the masses could be overestimated, at least to some extent \citep{2025arXiv250316595R, 2025arXiv251000103T}. The discrepancy between the local and high-redshift $M_\star-m_\bullet$ relation could then possibly be attributed to both selection biases and measurement uncertainties  \citep{2025ApJ...981...19L}. 

The fact that our simulated galaxies match the local $M_\star-m_\bullet$ relation is not surprising, as the BH accretion and feedback models self-regulate SMBH growth and are calibrated to reproduce the local $M_\star-m_\bullet$ relation at low redshifts and similar evolution on the $M_\star-m_\bullet$ plane is seen in many other cosmological simulations \citep{Habouzit2021}. The growth of SMBHs in galaxies is delayed when switching from repositioning to a dynamical friction subgrid model, where the SMBH can wander around the centre of the system until it has reached a sufficiently large mass. One possible solution for creating more massive SMBHs in the early Universe is to allow for super--Eddington accretion. If super--Eddington accretion is sustainable for multiple Myr, SMBHs would be able to grow above the local $M_\star-m_\bullet$ relation (see e.g. \citealt{Lupi2024}). 

The choice of seeding model can assist in the growth of SMBHs at high redshift. For example \citet{2024MNRAS.533.1907B} show that the SMBH mass growth is dominated by mergers in the early Universe when using a seeding criteria based on gas properties. Interestingly, we do not see a similar effect, as seven of the nine SMBHs have only grown by gas accretion before reaching the $m_\mathrm{Ketju}$ mass limit. The largest driver for the disparity is most likely the fact that we employ a larger halo mass limit for BH seeding 
than the galaxy mass criterion used in \citet{2024MNRAS.533.1907B}. This highlights the importance of a sophisticated seeding model if the evolution of BH seeds at very high redshifts is studied.

Stuyding the size evolution of the stellar component (Figure \ref{fig:size-evolution}) we saw a phase of compaction, starting roughly when the stellar mass reached $10^9\,\mathrm{M}_\odot$. A similar phase of compaction is also seen in other simulations \citep{Zolotov2015, 2023MNRAS.522.4515L}, although for galaxies with higher stellar masses ($\gtrsim 10^{10}\ \mathrm{M}_\odot$) and at lower redshifts ($z\sim 2-4$). These studies report a wet compaction, where gas falls inward due to mergers, disc instabilities or cold streams, which produces a `blue nugget' galaxy. The compaction of those systems stop as gas starts to be depleted from the centre, resulting in quenched central regions. 

Recently, \citet{cataldi2025galaxysizescompactnesscosmic} reported a phase of compaction for galaxies in the \textsc{FirstLight} simulation suite, which have a very high mass resolution ($m_\mathrm{dm}=10^4\ \mathrm{M}_\odot$). Our results are in agreement, as the compactions occur in the same redshift range, in galaxies with similar masses and sizes. It is important to note that not all simulations see such a compaction phase (e.g. \citealt{2025MNRAS.544.1732M}). \textsc{FirstLight} does not include AGN feedback, meaning that although the SMBHs in our simulation start to grow in mass around the same time that the compaction ends, SMBH feedback most likely does not strongly affect the size growth for the compact galaxies. Interestingly, the end of compaction in \textsc{FirstLight} differs from the lower redshift studies of \citet{Zolotov2015} and \citet{2023MNRAS.522.4515L}. The central regions of the galaxies are not fully depleted of gas and thus not entirely quenched. Rather, star formation continues in the central region, but the gas in the outer region also fulfil the conditions for star formation, thus resulting in an increase in the half-mass radius. 

In this aspect, our simulations are different, as the gas fraction in the central regions for most of our galaxies 
drop to a very small value of $\lesssim 1\%$ (see Figure \ref{fig:multipanel}). 
This difference might be due to differences in the feedback implementation, as our study includes BH feedback. 
The SMBH feedback pushes gas away from the central region, and together with the centrally concentrated star formation results in central regions depleted in gas. Therefore, although simulated galaxies do increase in size without SMBHs, the inclusion of SMBHs is still important for the gas content in the central regions of galaxies.

Interestingly, the sizes of the compact simulated galaxies agree well with JWST observations of LRDs. This suggests that a LRD-like phase could be a common phase in early galaxy evolution, at least in terms of galaxy sizes. Note that we here only simulate one zoom-in region and only analyse galaxies in which the SMBH masses have reached $m_\mathrm{Ketju}=10^7\,\mathrm{M}_\odot/h$. A study in a larger volume also focusing on systems with lower mass SMBHs would be needed to study whether all galaxies have a similar early evolution phase.

For SMBHs, it has been suggested that LRDs are both the birthplaces of heavy seeds \citep{2025ApJ...994...40P} and that they contain SMBHs of masses $\sim 10^7-10^8\ \mathrm{M}_\odot$ \citep{2024ApJ...964...39G}. However, it is also possible that the light from LRDs is dominated by star formation \citep{Baggen2024}. In our case, the galaxies matching the sizes and masses of LRDs also include efficiently accreting SMBHs, although the SEDs of the two shown galaxies (Figures \ref{fig:SKIRT_galaxyA} and \ref{fig:SKIRT_galaxyB}) are mostly dominated by the stellar component in the relatively narrow wavelength range shown in the figures. In fact, SMBH growth is inefficient before galaxies reach the masses and sizes matching observations of LRDs (see Figures \ref{fig:size-evolution} and \ref{fig:multipanel}), akin to the suggestion that the LRDs host the first AGN events of SMBH accretion \citep{2025ApJ...988L..22I}. 

It is still uncertain what type of systems LRDs will evolve into. It has been suggested that LRDs will acquire an extended stellar component over time \citep{2025arXiv250704011B}, with cold gas accretion as one of the proposed mechanisms. We see a similar type of evolution, as galaxy stellar sizes grow due to off-central star formation, maintained by a continuous inflow of gas. As our simulation was stopped at redshift $z=5$, the sizes and masses began to match with those of cMQGs observed in the redshift range $3<z<4$ \citep{Kawinwanichakij25}. The simulated galaxies have already depleted the central regions from gas, and other simulations reporting a phase of compaction of higher mass galaxies also have galaxies quenching after the phase of compaction ends \citep{Zolotov2015}. If our simulated galaxies become fully quenched, it could link LRD-like systems as the progenitors of cMQGs. 

The star formation in our simulation is heavily dominated by in-situ star formation with nearly all stars being formed in-situ at $z_\mathrm{compact}$ (\autoref{table:properties}). The in-situ dominated star formation at high redshifts also fits into the picture of two-phase galaxy evolution (e.g. \citealt{2009ApJ...699L.178N,2010ApJ...725.2312O,2012ApJ...754..115J}), where the growth of low redshift massive early-type galaxies was dominated by in-situ star formation at high redshifts. For lower mass systems, the mass growth of the stellar component remains dominated by in-situ star formation throughout the cosmic epoch \citep{2016MNRAS.458.2371R}.

\subsection{SMBH Dynamics in a cosmological context}
\label{sect:discussion_dynamics}

The zoom-in simulation presented here including \ketju{} is the first time that SMBHs have been modelled from the moment of seeding all the way down to their GW driven coalescence at sub-pc scale separations in one single simulation. Previously, in order to evolve binaries formed in a cosmological environment to sub-pc scale separations, either re-simulations with higher resolution (e.g. \citealt{Fastidio2024, Chen2024_magicsI}) or hand-picking a  time at which regularised integration is switched on \citep{Mannerkoski2021, Mannerkoski2022} was required.

All three \ketju{} integrated mergers coalesce in less than $50\,\mathrm{Myr}$ after becoming gravitationally bound, suggesting that mergers of SMBHs at high redshifts occur very rapidly, at least in compact massive galaxies. The rapid binary hardening is driven by very high stellar densities. The central densities are a result of a centrally focused star formation and other works have highlighted how this helps SMBH binaries to coalesce \citep{Liao_rabbits_I_2024}. Both the BH merger timescales and stellar densities are similar to the simulation run by \citet{Khan2016}. Their simulations are re-runs of systems chosen from the Argo simulation \citep{2015MNRAS.446.1939F}, starting at redshift $z\approx 3.5$. The final orbital evolution of the SMBH binary is modelled using a $N$-body code $\phi$-GPU \citep{phiGPU} by extracting the central region from the simulations and turning the gas particles into stellar particles. Although their method of generating initial conditions for a SMBH binary and galaxy merger clearly differs from our approach, the central densities of the galaxy merger presented in \citet{Khan2016} are similar to the stellar densities of the galaxies in our simulation (Figure \ref{fig:3d_densities}). Importantly, the SMBH binary in \citet{Khan2016} reaches coalescence in $\sim 10\,\mathrm{Myr}$ after the binary reaches a separation of $100\,\mathrm{pc}$, a timescale that matches our binary evolution (Figures \ref{fig:zoom-in_binary_params} and \ref{fig:binary_properties}). The agreement in merger timescales further demonstrates that high central stellar densities drive SMBH binaries to coalescence on short timescales.

Since the rapid mergers are due to large stellar densities in the central regions, precise modelling of star formation and stellar physics at high redshifts is important when studying the dynamics of SMBH binaries. Although galaxies are expected to be more centrally concentrated in the early Universe in terms of the size of the stellar component (e.g. \citealt{2012ApJ...756L..12M, 2024ApJ...963....9M}), extending the study to a larger variety of galaxies and to even higher redshifts would extend our knowledge about the evolution of SMBH binaries in the early Universe. Such simulations would require increased mass resolution compared to the study in this paper and an improved SMBH seeding prescription (e.g. \citealt{2025MNRAS.542.2597C})

Combining the previous results of the duration of SMBH binaries acquired with \ketju{} in cosmological zoom-in simulations (\citealt{Mannerkoski2022}, their Figure 3) together with the binaries presented in Figure \ref{fig:zoom-in_binary_params}, we see that the duration of the SMBH merging process from the initial dynamical friction driven phase down to the final coalescence can vary from a few Myr up to even a Gyr. The dominating effects for binary evolution at small separations (three-body scattering and GW emission) are often added to simulated mergers in post-processing. Recent examples of this include e.g. \cite{Bellovary2024} for SMBH mergers resulting from dwarf galaxies merging with Milky Way-like galaxies in zoom-in simulations, and \cite{Astrid_GWB} for the SMBH merger population in the ASTRID simulation. The former added a maximum of $1\,\mathrm{Gyr}$ to the duration of inspirals, while the latter chose a binary lifetime of $\tau_f=500\,\mathrm{Myr}$ for the entire population. Our results suggest that post-processing prescriptions should allow a large range of merger delays (see e.g. \citealt{Kelley2017}).

The optimal use for the new combined model of \ketju{} and the dynamical friction subgrid model would be on larger simulation volumes. Such simulations would include SMBH binaries with various mass ratios in diverse environments and would, for example, allow us to study the eccentricity evolution of SMBH binaries in frequency wavebands which current and future GW observation missions target. Such a study would especially be interesting as we showed that \ketju{} is able to model binaries through the PTA waveband (Figure \ref{fig:gw-strain}) and a larger binary population would provide insight to whether all high redshift binaries reach coalescence on a timescale shorter than $100\ \mathrm{Myr}$ after becoming bound.  In addition, running a simulation to a lower redshift could be used to study the effect which SMBH binaries have on their host galaxies (see e.g. \citealt{Rantala_nuggets} for a such a study a non-cosmological setting). 

Another interesting future aspect would be to study the mergers of less massive SMBHs. Improving the mass resolution by an order of magnitude would allow us to use this model with \ketju{} being enabled for SMBHs with $m_\bullet\gtrsim 10^6\,\mathrm{M}_\odot$. Binaries of such SMBHs would have a larger fraction of their evolution spent in the LISA waveband compared to the larger mass binaries (see e.g. \citealt{Liao2025}). The decreased mass would be of the same order as the seed mass used in this study, so higher mass resolution would allow \ketju{} to be used for all SMBHs, or alternatively, the seed mass could be decreased.
Various studies have discussed the difficulty of sinking seed mass SMBHs (e.g. \citealt{Ma_seeds_2021, 2023MNRAS.525.1479D, Partmann_mergers}) and we showed that moving away from repositioning delays the beginning of SMBH accretion. Hence, it would be interesting to see how also the merger rate of seed mass SMBHs is affected.

\section{Conclusions}
\label{sect:conclusions}

In this study, we have for the first time modelled SMBHs in a cosmological setting from the moment of seeding to their GW-emission driven coalescence at sub-pc separations in a single simulation. This is made possible by combining the regularised integrator \ketju{} with a dynamical friction subgrid model. Previous simulations either stop at larger separations (\citealt{Romulus, Newhorizon, astrid}), or restart simulations from a snapshot with an different code \citep{Mannerkoski2021, Mannerkoski2022}, or alternatively resimulate mergers in a non-cosmological setting with an altogether different code at an improved resolution (e.g \citealt{Khan2016, Chen2024_magicsI}). 

Before running the cosmological zoom-in simulations, we studied the dynamics of single SMBHs in two settings. 
We started by studying the sinking of a SMBH on a circular orbit in a Plummer sphere. Interestingly \textsc{ketju} managed to sink the SMBH into the centre relatively well even when the mass ratio between the SMBH and stellar particle mass was set to be $m_\bullet/m_\star=10$. SMBHs in isolated systems should stay at the centre and therefore we run a series of simulation with an SMBH located at the centre of the system without initial velocity. The Brownian motion of a SMBH in the centre of a Hernquist sphere increases notably when the SMBH mass was decreased to around $10^6\ \mathrm{M}_\odot$ ($m_\bullet/m_\star \approx 10$), while the dynamical friction subgrid model managed to keep the SMBH within a stellar softening length from the centre even when $m_\bullet/m_\star\leq 10$. \ketju{} keeps the SMBH closer to the centre than the subgrid model when the SMBH mass is $\gtrsim 10^7\ \mathrm{M}_\odot$ ($m_\bullet/m_\star \approx 100$). Based on these tests, we created a hybrid model of the two codes with the switch to \ketju{} occurring when the mass ratio $m_\bullet/m_\star\approx 120$ is reached.

From the evolution of galaxies, we saw that removing BH repositioning delays the beginning of the SMBH mass accretion due to seed mass SMBHs orbiting around the centre of the galaxy. As soon as SMBHs begin to grow in galaxies, the systems are on the local stellar mass--SMBH mass relation. Intriguingly, galaxies go through a phase of compaction as they grow with the compact masses and sizes agreeing with JWST observations of LRDs. The SEDs from two different simulated galaxies (galaxy A at $z=5$ and galaxy B at $z=7.99$) are mostly dominated by the stellar component in the observed wavelength band. We also note that the broadband colors of the simulated galaxies are in a general agreement with observed LRDs, although we stress that a more detailed comparison would require individual emission lines, not included in our simplified SEDs. The sizes of our simulated galaxies grow again after compaction ends due to off-centre star formation, which is fuelled by continuous gas inflows to the galaxies. Other studies have reported a similar phase at lower redshifts and for higher stellar mass systems (e.g \citealt{Zolotov2015}). 

The dynamics of SMBHs and the early evolution of galaxies was studied using cosmological zoom-in simulations run to redshift 5. Three versions of the simulation were run: one using the created hybrid of \ketju{} and the dynamical friction subgrid model, one using just the subgrid model and one using repositioning. The dynamics of SMBHs modelled with \ketju{} and the dynamical friction subgrid model agree with each other until the criteria for merger with the subgrid model is reached. The \ketju{} integrated binaries merge on timescales less than $50\ \mathrm{Myr}$ after becoming bound with the main cause being the large central densities of the host galaxies. The simulated binaries are evolved through the PTA wavebands with \ketju{}, down to a frequency range where a phenomenological code PhenomD can be used to model the very final orbits of a SMBH binary. Importantly, we find that the SMBH binaries from the JWST-observable first galaxies where SMBHs efficiently grow can be observed with LISA.

The created hybrid model for BH dynamics is an important step for SMBH modelling in a cosmological environments. The zoom-in simulations performed here demonstrated the general capability of modelling SMBH binaries into the GW emission regime. The combination of \ketju{} with a dynamical friction subgrid model provides the opportunity to use \ketju{} for both a larger redshift range and for larger cosmological volumes.

\section*{Acknowledgements}

A.K., P.H.J., A.R., T.T and B.R acknowledge the support by the European Research Council via ERC Consolidator grant KETJU (no. 818930). P.H.J. and T.T. also acknowledge the support of the Research
Council of Finland grant 339127. T.N. acknowledges support from the Deutsche Forschungsgemeinschaft (DFG, German Research Foundation) under Germany’s Excellence Strategy--
EXC--2094--390783311 from the DFG Cluster of Excellence “ORIGINS.” 
S.L. acknowledges the support by the National Natural Science Foundation of China (NSFC) grant (no. 12588202, 12473015).

We list here the roles and contributions of the authors
according to the Contributor Roles Taxonomy (CRediT).\footnote{\href{https://credit.niso.org/}{https://credit.niso.org/}} A.K.: Conceptualization,
Investigation, Formal analysis, Data curation, Writing -- original draft. P.H.J.: Conceptualization, Supervision, Writing -- original draft. A.R.: Writing -- review \& editing. T.T.: Formal analysis, Writing -- original draft. A.R.: Writing -- review \& editing. T.N.: Conceptualization, Writing -- review \& editing. S.L.: Writing -- review \& editing. B.R.: Writing -- review \& editing.

\textit{Software:} \ketju{} \citep{2017ApJ...840...53R}, \textsc{gadget-3} \citep{2005MNRAS.364.1105S}, NumPy \citep{2020Natur.585..357H}, SciPy \citep{2020NatMe..17..261V}, Matplotlib \citep{2007CSE.....9...90H}, pygad \citep{2020MNRAS.496..152R}, BOWIE \citep{2019MNRAS.483.3108K} and Hasasia \citep{Hazboun2019Hasasia}.

\section*{Data Availability}

Data from all presented simulations can be made available upon a reasonable request.



\bibliographystyle{mnras}
\bibliography{refs}




\appendix

\section{Differences in large-scales between the simulations}
\label{sect:run_diffs}

\begin{figure*}
    \centering
    \includegraphics[width=\linewidth]{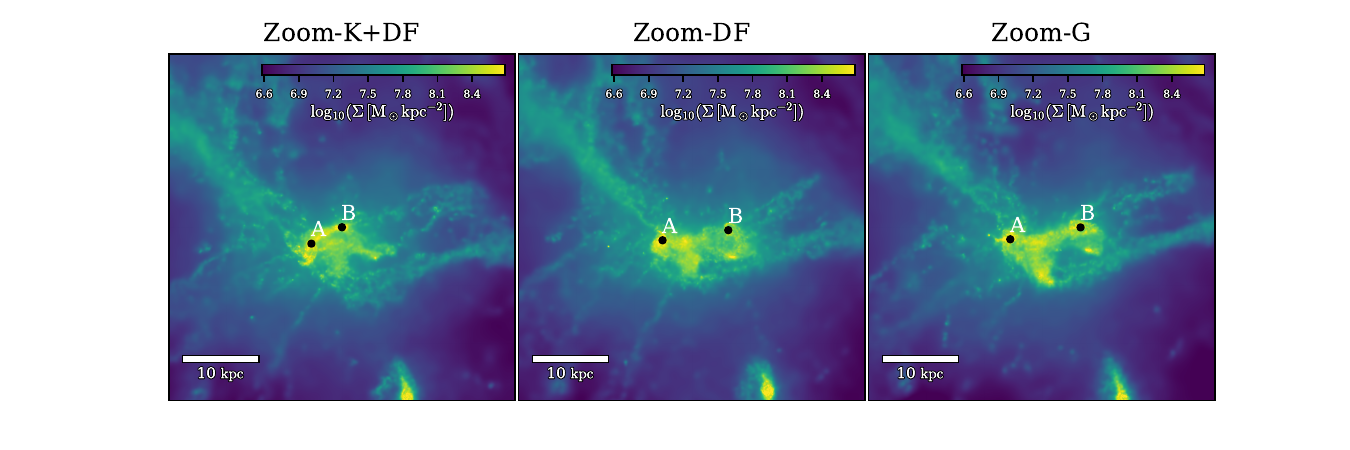}
    \caption{Snapshot showing the gas surface density of galaxies A and B and their surrounding at a redshift of $z=5.27$. From left to right, the panels show the zoom-K+DF, zoom-DF and zoom-G version of the simulated volume. The black points represent the positions of the SMBHs.
    }
    \label{fig:run_diff}
\end{figure*}

In Figure \ref{fig:run_diff} we show the surface density of gas from a region containing galaxies A and B at a redshift of $z=5.27$, around $20\ \mathrm{Myr}$ before the two SMBHs coalesce in the simulation where \ketju{} was enabled. The three panels from left to right show the simulations run with \ketju{} combined with the dynamical friction model (zoom-K+DF), the dynamical friction model only (zoom-DF) and the \textsc{gadget-3} run with repositioning (zoom-G), respectively. 
While all three simulations start from identical initial conditions, they evolve independently. 

The most noticeable difference is in the position of galaxy B, which is considerably closer to galaxy A in the simulation zoom-K+DF compared to the other two simulations. The position of galaxy A also slightly changes. These differences on global galactic scales also causes the galaxies A and B to merge earlier in the simulation zoom-K+DF, resulting also in an earlier SMBH coalescence. There are also visible differences in the positions of smaller structures as all three simulations have unique gas surface densities. It is important to note that the merger D$\rightarrow$A has already occurred before the shown snapshot, affecting the large-scale structures of the galaxies. 

It is also well known that properties of halo substructures, such as their positions, are subject to change because of multiple numerical effects, including numerical resolution \citep{2004MNRAS.355..819G}, small changes to initial conditions \citep{2008MNRAS.387..397T} and the choice of code and integration parameters \citep{1999ApJ...525..554F}. Therefore it is not too unexpected that the location of galaxies begins to deviate between the simulations with different SMBH dynamics modelling schemes.

\section{Characteristic strain calculation}
\label{sect:gw_calc}
Following \citet{Amaro-Seoane2010} and \citet{Kelley2017}, the GW strain amplitude $h_s$ of an individual source can be calculated as a sum of strain amplitudes $h_{s,n}$ at different harmonics $n$. The total strain amplitude at orbital frequency $f_r$ is then
\begin{equation}
    h_s^2(f_r) = \left.\sum_{n=1}^\infty h_{s,n}^2(f_h)\left(\frac{2}{n}\right)^2 \right|_{f_h=f_r/n}.
\end{equation}
We use $n_\mathrm{max}=250$ harmonics in the GW calculations. We check that the inclusion of higher harmonics has a negligible effect on the results. The characteristic strain $h_c$ can be obtained via
\begin{equation}
    h_c(f) = \sqrt{N(f)}h_s(f),
\end{equation}
where $N$ is the number of cycles spent around frequency $f$ on an interval $\Delta f\approx f$. Therefore the characteristic strain is
\begin{equation}
    h_c^2(f_r) = \left.\sum_{n=1}^\infty \left(\sqrt{N}h_{s,n}(f_h)\right)^2\left(\frac{2}{n}\right)^2 \right|_{f_h=f_r/n},
\end{equation}
where the characteristic strain at a single harmonic $n$ is \citep{Berentzen2009}
\begin{equation}
\begin{split}
    h_{c,n}(f) & =\sqrt{N(f)}h_{s,n}(f) \\
    & = \frac{37\sqrt{5}}{64\sqrt{6}\pi^{2/3}}\frac{(G\mathcal{M})^{5/6}}{c^{3/2}d(z)}\sqrt{g(n,e)}f^{-1/6},
\end{split}
\label{eq:strain}
\end{equation}
with $\mathcal{M}$ being the chirp mass of the binary, $d(z)$ the comoving distance from the observer and 
\begin{equation}
\begin{split}
    g(n,e)= & \frac{n^4}{32}\bigg[ \bigg(J_{n-2}(ne)- 2eJ_{n-1}(ne) \\
    & +\frac{2}{n}J_n(ne)+2eJ_{n+1}-J_{n+2}(ne) \bigg)^2 \\
    & + (1-e)^2\left(J_{n-2}(ne)-2J_n(ne)+J_{n+2}(ne) \right)^2 \\
    & + \frac{4}{3n^2}J_n^2(ne) \bigg]
\end{split}
\end{equation}
is a function describing the relative contribution of radiated power in harmonic $n$ \citep{PetersMathews}, where $J_n$ is the Bessel function of the first kind or order $n$. Note that we use equation \ref{eq:strain} with $f_h=f_r/n$, while \cite{Berentzen2009} use $f_\mathrm{obs}$. For the equation to be true, we assume that $N\lesssim f_\mathrm{obs}\tau$, where $\tau$ is the observation time. In the case of $N\gtrsim f_\mathrm{obs}\tau$, the characteristic strain at a single harmonic is instead \citep{Berentzen2009}
\begin{equation}
    h_{c,n}(f)  = \frac{37\pi^{2/3}}{16}\frac{(G\mathcal{M})^{5/3}}{c^{4}d(z)}\sqrt{\tau g(n,e)}f^{7/6}.
\end{equation}
The output of \ketju{} is saved every $5\times10^4\,\mathrm{yr}$ and the final output exists  $\sim 15\,000-35\,000\,\mathrm{yr}$ before the integrated coalescence for the three binaries. In order to include the integrated but not saved evolution in the characteristic strain calculation, the binaries are evolved from the final saved output using the 2.5PN term \citep{Peters64}
\begin{equation}
    \frac{\mathrm{d} a}{\mathrm{d}t}=-\frac{64}{5}\frac{G^3(m_1+m_2)m_1m_2}{c^5a^3(1-e^2)^{7/2}}\left(1+\frac{73}{24}e^2+\frac{37}{96}e^4 \right)
\end{equation}
and 
\begin{equation}
    \frac{\mathrm{d} e}{\mathrm{d}t}=-\frac{304}{15}\frac{G^3(m_1+m_2)m_1m_2}{c^5a^4(1-e^2)^{5/2}}\left(1+\frac{121}{304}e^2 \right)e.
\end{equation}


\bsp	
\label{lastpage}
\end{document}